# Mercury (Hg) in meteorites: variations in abundance, thermal release profile, mass-dependent and mass-independent isotopic fractionation

Matthias M. M. Meier[1,2], Christophe Cloquet[1], Bernard Marty[1]

[1]Centre de Recherches Pétrographiques et Géochimiques (CRPG), UMR 7358, Université de Lorraine, CNRS, 54500 Vandœuvre-lès-Nancy, France.
[2]Institute of Geochemistry and Petrology, ETH Zurich, Clausiusstrasse 25, 8092 Zurich, Switzerland.

**Abstract** – *We have measured the concentration, isotopic composition and thermal release profiles of Mercury (Hg) in a suite of meteorites, including both chondrites and achondrites. We find large variations in Hg concentration between different meteorites (ca. 10 ppb to 14'000 ppb), with the highest concentration orders of magnitude above the expected bulk solar system silicates value. From the presence of several different Hg carrier phases in thermal release profiles (150 – 650 °C), we argue that these variations are unlikely to be mainly due to terrestrial contamination. The Hg abundance of meteorites shows no correlation with petrographic type, or mass-dependent fractionation of Hg isotopes. Most carbonaceous chondrites show mass-independent enrichments in the odd-numbered isotopes $^{199}$Hg and $^{201}$Hg. We show that the enrichments are not nucleosynthetic, as we do not find corresponding nucleosynthetic deficits of $^{196}$Hg. Instead, they can partially be explained by Hg evaporation and redeposition during heating of asteroids from primordial radionuclides and late-stage impact heating. Non-carbonaceous chondrites, most achondrites and the Earth do not show these enrichments in vapor-phase Hg. All meteorites studied here have however isotopically light Hg ($\delta^{202}$Hg = ~-7 to -1) relative to the Earth's average crustal values, which could suggest that the Earth has lost a significant fraction of its primordial Hg. However, the late accretion of carbonaceous chondritic material on the order of ~2%, which has been suggested to account for the water, carbon, nitrogen and noble gas inventories of the Earth, can also contribute most or all of the Earth's current Hg budget. In this case, the isotopically heavy Hg of the Earth's crust would have to be the result of isotopic fractionation between surface and deep-Earth reservoirs.*

# 1. Introduction

Mercury (Hg) is an element of high cosmochemical interest, not least due to is unusual properties. Like a noble gas, it is highly volatile and comparatively rare in silicate matter, with abundances typically measured in parts per billion (ppb or ng/g). It has a high vapor pressure for a metal, usually forms a mono-atomic gas ($Hg_{(0)}$) in the vapor phase, and does not easily enter chemical bonds. In contrast to noble gases, however, Hg has a strong affinity to organic matter. Mercury is the heaviest known element with more than three stable isotopes, and thus the heaviest element where internal normalization of isotope ratios can be used. Its seven stable isotopes (196, 198-202, 204) span a mass range ($\Delta M/M$) of ~4%. Both mass-dependent and mass-independent isotope fractionation processes (MDF, MIF) have been confirmed in a variety of terrestrial environments (see below). The seven isotopes of Hg have different nucleosynthetic origins: the lightest and least abundant (~0.15%) isotope, [196]Hg, is produced exclusively by the p-process, while the other six isotopes have varying contributions from the s- and r-processes (Palme & Beer, 1993; Jaschek & Jaschek, 1995, Arlandini et al., 1999). Nucleosynthetic models predict [196]Hg/[198]Hg excesses on the order of several 1000‰ for supernova (SN) ejecta (e.g., Rauscher et al., 2002). Therefore, even a small fraction of SN-derived matter injected at a late stage into the solar nebula, similar to what has been suggested for short-lived radionuclides (e.g., Boss & Keiser, 2012), should lead to measurable [196]Hg excesses, at least if the SN-derived Hg escapes complete re-mixing with nebular Hg before incorporation into silicates. The high volatility of Hg further makes it a potential tracer for temperature-sensitive processes during all stages of planet formation: evaporation and re-condensation in the solar nebula, parent-body heating, de-volatilization and alteration, volatile loss during giant impacts (for planets), and late delivery of volatile elements to Earth and the other terrestrial planets (e.g., the "late veneer").

Given this potential, it is surprising how little is actually known about the abundance and isotopic composition of Hg in meteorites, despite more than half a century of study. Even the solar system abundance remains unclear since Hg cannot be detected in the solar photosphere (Grevesse, 1970; Grevesse et al., 2015). The Hg content in CI chondrites (Ivuna-type; e.g. Orgueil), which has in most cases been measured by either neutron activation analysis (NAA) or wet chemistry, is of little help in that respect, as it varies by up to three orders of magnitude between different CI chondrites, and even between individual samples of the same CI chondrite (Lauretta et al., 1999). From nuclear reaction cross-sections, and the known abundance of elements with similar masses, Beer & Macklin (1985) calculate a Hg abundance of $0.34 \times 10^{-6}$ atoms of Hg per atom of Si (given by these authors without uncertainty), which corresponds to an "expected" concentration of 260 ppb Hg for CI chondrites (Lauretta et al., 1999). It is currently not clear how Hg was incorporated into meteorites, i.e., we do not know whether we should indeed expect 260 ppb Hg in CI chondrites. Lauretta et al. (1999) suggested a model of chemisorption of Hg onto Fe-grain surfaces in the solar nebula, which is capable, for certain temperatures and Fe-grain sizes, to condense the full solar nebula content of Hg. In the following, we will simply assume that the solar abundance of Hg was initially fully condensed into meteorites. We will call this initial abundance of 260 ppb Hg the "bulk solar system silicates value", or $Hg_{BS3}$.

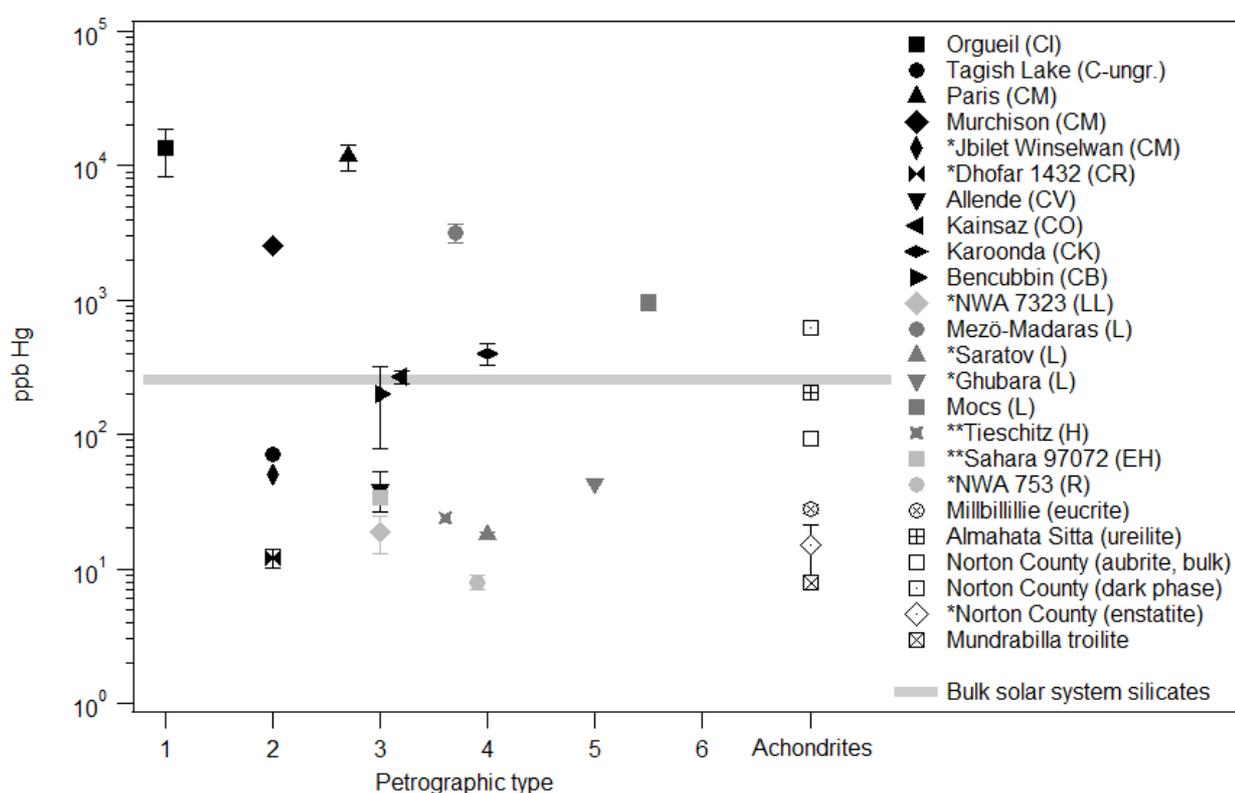

*Figure 1: Abundance of Hg (ordinate; logarithmic) vs. petrographic type (abscissa). *No isotopic analysis. **Isotopic analysis restricted to mass-dependent fractionation. The gray horizontal line corresponds to the estimated bulk solar system abundance of Hg (HgBS3) calculated from the abundance of neighboring elements (260 ppb; Lauretta et al., 1999).*

Meteorites with abundances significantly above the $Hg_{BS3}$ value would thus have experienced some process of additional Hg acquisition, while lower values require either Hg loss or inefficient initial acquisition. Whether these processes happened on Earth (terrestrial "contamination" and "loss") or in space (e.g., through thermal processes on the parent body) is unclear. Some authors have dismissed the exceptionally high Hg abundances measured in Orgueil (up to 500'000 ppb) as due to terrestrial contamination (e.g., Palme & Beer, 1993). For the majority of the meteorites analyzed so far, the Hg concentration was measured using NAA, which can detect only two of the seven isotopes, $^{196}$Hg and $^{202}$Hg. Strong variations in the $^{196}$Hg/$^{202}$Hg ratio (of up to a few 1000‰, both positive and negative) have been reported in many early NAA studies (see summary by Lauretta et al., 1999). However, Ebihara et al. (1998), Lauretta et al. (1999) and Kumar et al. (2001) have suggested that interference from $^{75}$Sn and $^{203}$Hg might have compromised the Hg signal in NAA studies, leading to both an overestimation of the total Hg content, and spurious isotopic variation in the measured $^{196}$Hg/$^{202}$Hg ratio. Lauretta et al. (2001) analyzed the carbonaceous chondrites Murchison (Mighei type; CM) and Allende (Vigarano type; CV), for which large isotopic variations in $^{196}$Hg/$^{202}$Hg had previously been reported in NAA studies. They used single- and multi-collector inductive-coupled plasma mass spectrometry (SC-ICP-MS and MC-ICP-MS, respectively) to find that the Hg isotopic composition of these meteorites was compatible with terrestrial Hg within 0.2-0.5‰, except for a small unexplained deficit in $^{200}$Hg in Murchison. Lauretta et al. (2001) also reported a Hg concentration of 294±15 ppb and 30.0±1.5 ppb Hg for Murchison and Allende, respectively, which is within the lower range of values determined by NAA for these meteorites.

The development of cold vapor (CV) MC-ICP-MS (Klaue et al., 2000), and the pioneering work by Lauretta et al. (1999; 2001), have inspired a large number of ICP-MS studies of Hg in Earth- and environmental science (Blum et al., 2014). Mass-independent isotopic fractionations (MIF) of Hg isotopes have since repeatedly been discovered in terrestrial environments, with both the even- and odd-numbered Hg isotopes affected (e.g., Chen et al., 2012; Demers et al., 2013). These MIFs allow the identification and quantification of different sources and processes contributing to the Hg measured in terrestrial samples, e.g., atmospheric, industrial and urban Hg (Estrade et al., 2010) or Hg from compact fluorescent lamps (Mead et al., 2013). Strong MIF of up to several permil have been observed in such unlikely seeming reservoirs as freshwater and marine fish (e.g. Bergquist & Blum, 2007), and Arctic snow (Sherman et al., 2010). A recent review of terrestrial Hg measurements and the processes thought to be responsible for the observed MIFs is given by Blum et al., 2014.

There are two processes inducing MIFs in the odd-numbered Hg isotopes ($^{199}$Hg, $^{201}$Hg): The "magnetic isotope effect" (MIE), which is due to the odd-numbered isotopes carrying a magnetic moment, resulting in a slightly different behavior during chemical reactions involving radicals; and the "nuclear volume effect" (NVE), where the slightly different ion radii of the odd-numbered isotopes are thought to result in isotopic fractionation during evaporation and condensation, such that the Hg vapor phase is enriched both mass-dependently in the light isotopes and mass-independently in the odd isotopes, with a $\Delta^{199}$Hg/$\delta^{202}$Hg slope of $\sim$-0.1 (Ghosh et al., 2013). Here, $\delta^{xxx}$Hg = $((({}^{xxx}$Hg/$^{198}$Hg$_{measured}$)/$({}^{xxx}$Hg/$^{198}$Hg$_{NIST\text{-}3133\ standard}$)) $-$ 1) $\times$ 1000, i.e., the deviation of an isotopic ratio in a sample relative to the NIST-3133 Hg standard, in permil (‰). The $\Delta^{xxx}$Hg values are deviations of the $\delta^{xxx}$Hg values from the mass-dependent fractionation value implied by $\delta^{202}$Hg, and defined as $\Delta^{xxx}$Hg = $\delta^{xxx}$Hg $-$ $\alpha \times \delta^{202}$Hg, with $\alpha$ = $(1/m_{198} - 1/m_{xxx})$ / $(1/m_{198} - 1/m_{202})$, and $m_{xxx}$ the mass of the isotope $^{xxx}$Hg in atomic mass units. Estrade et al. (2009) performed Hg evaporation experiments under equilibrium (closed) and dynamic (open) conditions, and found MIF effects on the odd-numbered isotopes in both cases, resulting in characteristic $\Delta^{199}$Hg/$\Delta^{201}$Hg ratios of 2.0$\pm$0.6 and 1.2$\pm$0.4, for equilibrium and dynamic conditions, respectively. Ghosh et al. (2013) provide a more precise value of 1.59$\pm$0.05 for the $\Delta^{199}$Hg/$\Delta^{201}$Hg ratio under equilibrium (closed) conditions. In other words, the MIF of $^{199}$Hg relative to the MIF of $^{201}$Hg depends on the fractionating process, allowing, in principle, the identification of the process.

While the last decade has thus seen significant progress in the understanding of Hg in terrestrial environments, the non-detection of Hg isotopic anomalies in meteorites by Lauretta et al. (2001) seems to have forestalled interest in additional high-precision ICP-MS analyses of Hg in meteorites. Given the progress in Hg analysis and ICP-MS precision, and the new possibilities offered by MIFs, such a study now seems long overdue (Meier et al., 2015a; Meier et al., 2015b; Wiederhold & Schönbächler, 2015). In this work, we measured the abundance, thermal release profile, and isotopic composition of Hg in different meteorite samples including ordinary, carbonaceous, enstatite and Rumuruti chondrites, as well as some achondrites (eucrites, aubrites, ureilites, and troilite from an iron meteorite).

## 2. Samples and Methods

### 2.1. Meteorite sample selection

We initially restricted our analysis to meteorites from recent, well-documented and quickly recovered falls with large available masses, to minimize weathering, entry heating and potential contamination effects in the individual samples. Work by Lauretta et al. (1999; 2001) led us to expect Hg concentrations between a few and a few hundred ppb, which requires relatively large masses of up to several 10 g for isotopic analysis. After we found some meteorites with Hg concentrations of several 1000 ppb, we relaxed these constraints and included meteorites with smaller available masses as well. Sample sources are given in Table 1, and include the Academy of Sciences in Moscow (ASM), the Centre de Recherches Pétrographiques et Géochimiques in Nancy (CRPG), Lund University (LU), the National Museum of Natural History in Paris (MNHN), the Natural History Museum in Vienna (NHMV), the South Australian Museum in Adelaide (SAM), the University of Bern (UBE), the University of Bayreuth (UBA), the University of New Mexico in Alberquerque (UNMA) and the West Australian Museum in Perth (WAM). The rest of the samples were obtained commercially (COM1 and COM2; see Table 1 for details).

## 2.2. Concentrations and thermal release profiles measured with a Direct Mercury Analyzer

We used a Milestone© SRL Direct Mercury Analyzer (DMA-80) at SARM CRPG Nancy to determine the Hg concentration in all samples. This instrument has an effective stable detection limit of ~1 ng Hg (e.g., a Hg concentration of 2 ppb in a 500 mg sample). The excellent accuracy and precision of the DMA applied to geologic and environmental samples has recently been confirmed by the analysis of a large suite of terrestrial standards (Marie et al., 2015). Prior to analysis, the samples were crushed using different designs of mortars (depending on the hardness of the sample). The fragments were then further comminuted using a dynamic crusher (for 3 minutes), down to a grain size of a few 10 μm. We performed a series of tests to confirm that the blank contribution added by the dynamic crusher was below the detection limit of the DMA. This included the measurement of a terrestrial basaltic glass containing 10 ppb Hg, which was first completely degassed by heating it to 2200 °C in a hard vacuum, before comminution using the dynamic crusher. The glass was then analyzed in the DMA, and found not to contain any remaining Hg above detection limit. We also compared the measured Hg content of Orgueil dust passed through the dynamic crusher with Orgueil dust crushed very gently by hand. The resulting Hg concentrations were ~30% lower for the hand-crushed sample (~9000 ppb vs. ~13000 ppb), which is within the observed variability of Orgueil material. The difference between the two powders is also much larger than any plausible contribution from dynamic crushing itself as per our first experiments. We therefore conclude that dynamic crushing of the samples does not add or remove significant amounts of Hg from our samples prior to DMA analysis. In each DMA run, sample masses between 5 and 500 mg were used. All samples were heated to 650 °C for 3 minutes to release Hg, which is then amalgamated with Au inside the DMA. The amalgam is then re-heated to 900 °C, and quantified spectroscopically (see Marie et al., 2015, for a more detailed description). After analysis, all samples were passed through the DMA again to confirm that all the Hg had been extracted. Thermal release profiles were gained by sequentially heating the samples to 150, 200, 250, 300, 350, 400, 450 and 650°C for 3 minutes each. Between each temperature step, the sample was cooled down to ambient temperatures.

### 2.3. Isotopic analysis by mass spectrometry

The isotopic analysis was done on a Thermo Neptune+ multi-collector inductive-coupled plasma mass spectrometer (MC-ICP-MS) at CRPG CNRS Nancy, using a similar protocol as Estrade et al. (2009). The powder resulting from the dynamic crusher (< 3g in all cases) was reacted with concentrated (65%) $HNO_3$ in quartz tubes within a high pressure asher (HPA) heated to 300 °C at 120 bar during 3 hours. The resulting liquid was then extracted from the tubes, cooled to -20 °C in a freezer, and diluted to $HNO_3$ concentrations <32% to suppress potential Hg loss in acid vapors. We performed a series of tests (using the DMA) with Hg-rich terrestrial samples to ascertain that this extraction method would retrieve all Hg from the sample, and could not find Hg in the residue significantly above blank level. Also, the Hg concentration of Hg-rich (powdered) terrestrial samples and the nitric acid solution from the HPA extraction (again determined using the DMA) were identical within uncertainties. For MC-ICP-MS analysis, the acid solution from the HPA extraction was reacted with $SnCl_2$ using a small glass reactor (see Estrade et al., 2009 for details). The resulting gaseous $Hg_{(0)}$ was fed directly into the mass spectrometer using Ar carrier gas. Signals were typically within 10-20% of the values expected from DMA analysis. This should not be interpreted in the sense that either DMA analysis or HPA extraction have yields significantly <100%, as the typical variation between two samples of the same meteorite is on the same order of magnitude (see uncertainties in concentrations determined with the DMA in Table 1). Initially, we measured the five abundant Hg isotopes (198, 199, 200, 201, 202) and [203]Tl, [205]Tl to estimate the instrumental mass-dependent isotope fractionation (session I). After the latter was found to be reasonably constant, we also measured the less abundant Hg isotopes (196, 204; sessions II to V). It was not possible to measure all Hg isotopes and Tl at the same time, given the current mechanical limits to cup configuration in the instrument. Depending on availability, solutions were run at 2, 4, 5 or 10 ppb Hg level, with standards adjusted accordingly. We used a NIST-3133 standard for standard-sample-bracketing, UM-Almadén and two CRPG standards (F-65 and RL-24) for long-term reproducibility. As shown in Table 1, our average values for NIST-3133 are in excellent agreement with published values (Blum & Bergquist, 2007). While our nominal value for UM-Almadén is somewhat high, it is still compatible with the values determined by other groups (Mead et al., 2013; Yin et al., 2013; Chen et al., 2010) within uncertainty.

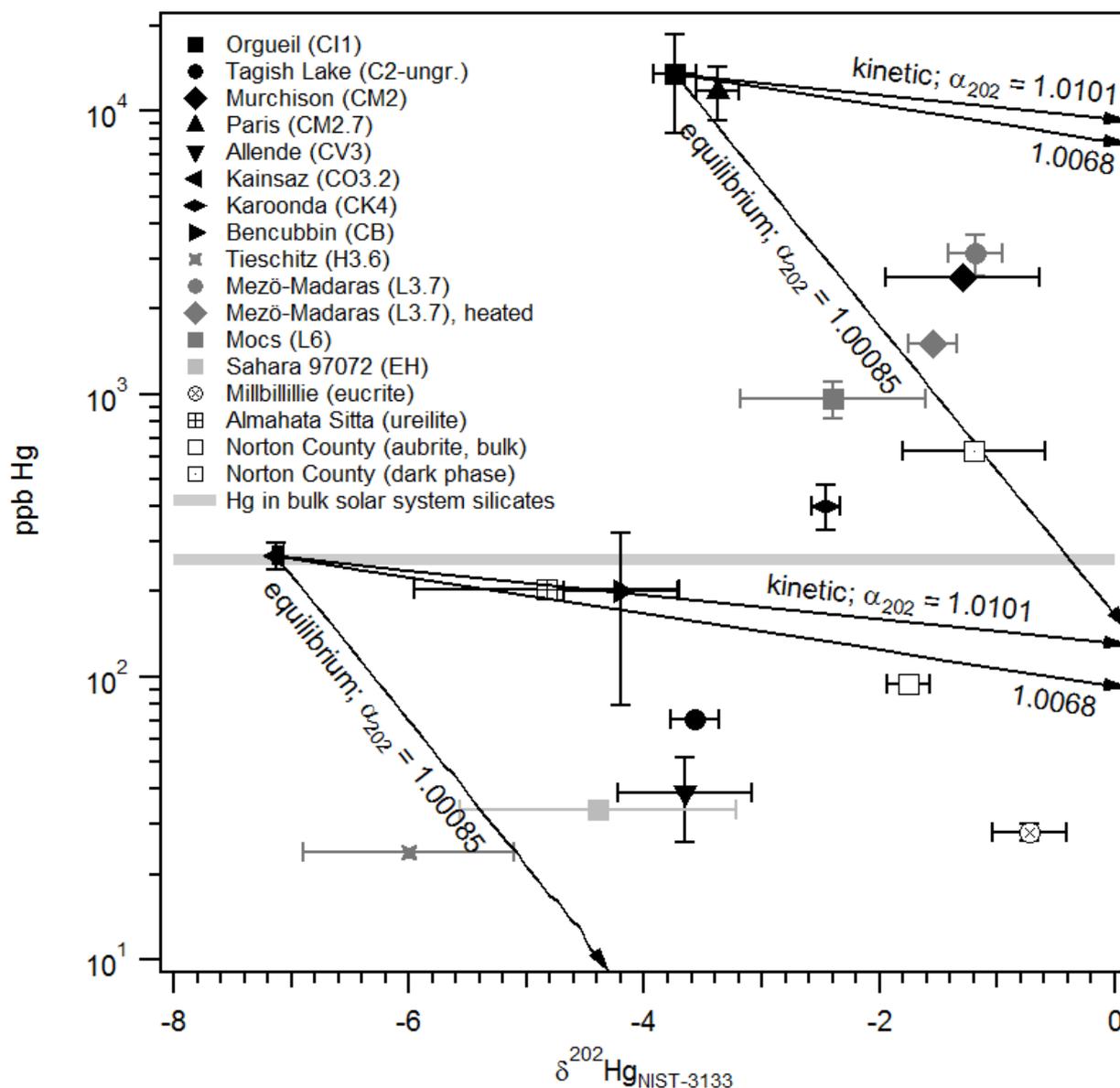

*Figure 2: Mercury abundance vs. mass-dependent isotope fractionation of Hg in different meteorites. All meteorite samples have isotopically lighter Hg than the Earth. The arrows starting at the data points of Orgueil (black square) and Kainsaz (black left-pointing triangle) represent fractionation lines under different conditions: theoretical kinetic (α202 = 1.0101), experimental kinetic (α202 = 1.0068; Estrade et al., 2009) and equilibrium (closed-system; α202 = 1.00086; Estrade et al., 2009). Note however that mass-dependent fractionation is temperature-dependent, with α202 = 1 for T >115 °C (i.e., vertical lines; Estrade et al., 2009).*

In accordance with the suggestion by Blum & Bergquist (2007), which has since been widely adopted in the Hg community, we use [198]Hg as our reference isotope, and normalize for MDF with [202]Hg. For samples where [200]Hg, [202]Hg and [204]Hg were measured, we calculated an average MDF from these three isotopes as well. The resulting MDF is within error of the MDF determined by normalizing to [202]Hg only. Only three of the seven Hg isotopes have potential isobaric interferences: [196]Hg and [198]Hg might be affected by interference from [196]Pt and [198]Pt, while [204]Hg might be affected by interference from [204]Pb. These interferences are however of little concern as the chemical-reduction method by which we extract the Hg from the liquid into the gaseous phase is Hg-specific. Nevertheless, given the very low relative abundance of [196]Hg and the high relative abundance of [196]Pt in some meteorites, we compared the peak-heights on masses 195 and 196 using the electron multiplier detector on the ICP-MS. We find a constant (i.e., sample-independent) Pt-background of ca. 300-500 counts on mass 195, which corresponds to 220-370 counts on mass 196. For a Hg concentration of 4 ppb, the lowest concentration for which we have measured [196]Hg, the Hg-peak on mass 196 is on the order of 8 mV or ~500'000 counts. Interference of background Pt would thus result in a [196]Hg excess of up to 0.7‰, which is however subtracted by the blank correction. Since Pt levels were not significantly elevated in samples vs. blanks or standards, we conclude that interference of Pt does not affect our measurement of [196]Hg.

## 3. Results

### 3.1. Terrestrial Hg contamination and loss

Due to the volatility of Hg and its affinity for organic matter, the potential of terrestrial Hg contamination in meteorites (in particular those rich in organic matter) has to be addressed seriously. Lauretta et al. (2001) subjected a sample of Allende to a saturated Hg atmosphere for 24 hours, but did not measure any difference in the Hg content of interior (unexposed) or exterior parts of the sample. The thermal release profiles of Hg from a fresh Allende sample, and a pre-heated ("baked") Allende sample exposed to a saturated Hg atmosphere for 24 hours were also clearly different with respect to peak shape and peak release temperature. The fresh Allende sample did not show any Hg at the release temperature of the contaminating Hg (ca. 190°C).

The meteorite with the highest Hg content in our study was Orgueil, at 14'000 ppb (Fig. 1). Similar or even higher Hg concentrations have been found in Orgueil by other authors (e.g., Palme & Beer, 1993; Wiederhold & Schönbächler, 2015). They have occasionally been suggested as being due to

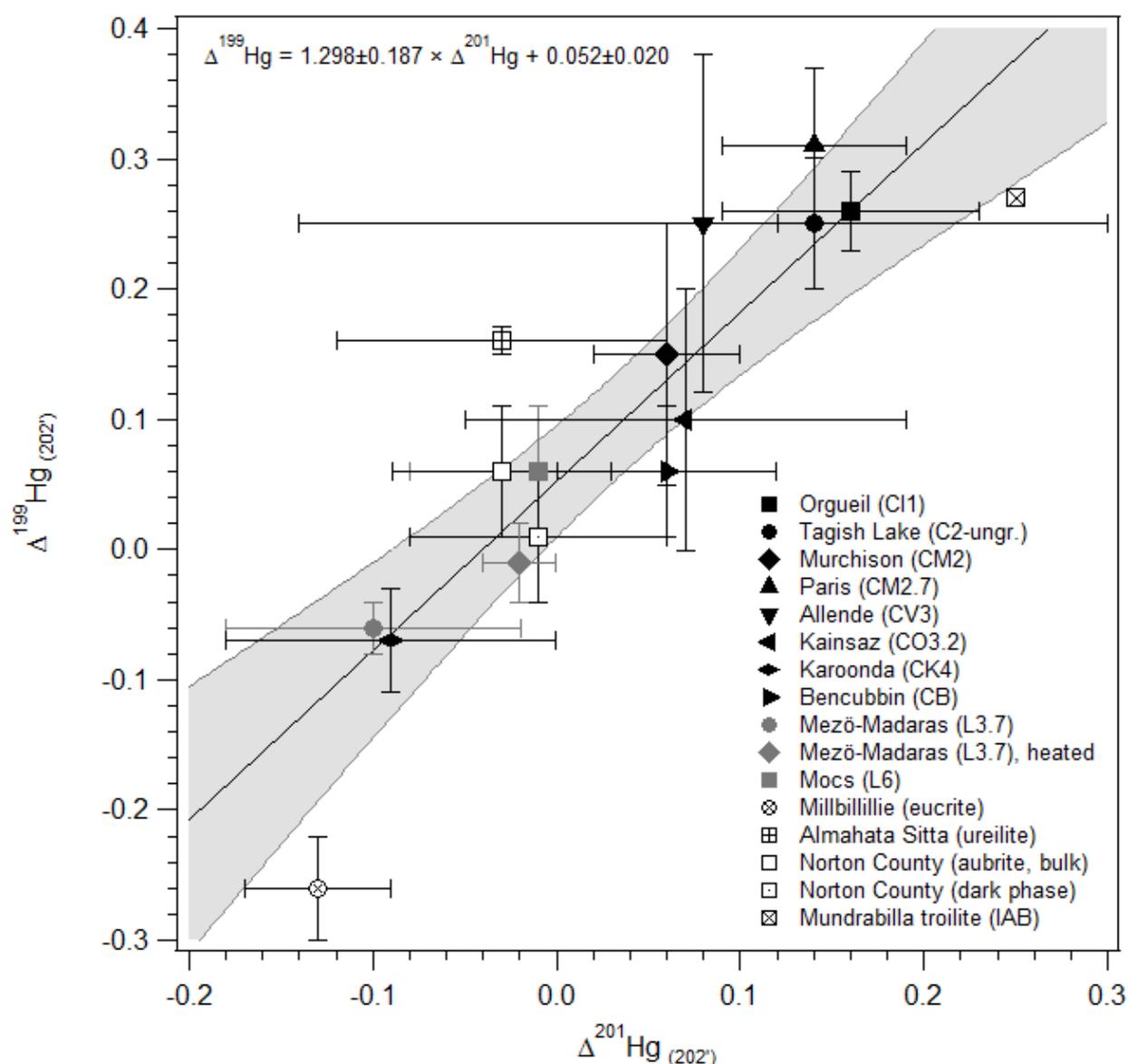

$$\Delta^{199}Hg = 1.298 \pm 0.187 \times \Delta^{201}Hg + 0.052 \pm 0.020$$

- ■ Orgueil (CI1)
- ● Tagish Lake (C2-ungr.)
- ◆ Murchison (CM2)
- ▲ Paris (CM2.7)
- ▼ Allende (CV3)
- ◀ Kainsaz (CO3.2)
- ◆ Karoonda (CK4)
- ▶ Bencubbin (CB)
- ● Mezö-Madaras (L3.7)
- ◆ Mezö-Madaras (L3.7), heated
- ■ Mocs (L6)
- ⊗ Millbillillie (eucrite)
- ⊞ Almahata Sitta (ureilite)
- □ Norton County (aubrite, bulk)
- ⊡ Norton County (dark phase)
- ⊠ Mundrabilla troilite (IAB)

*Figure 3: Co-variation of odd-numbered Hg isotope excesses and deficits. Within error, all meteorites plot on a single trend line with a slope of 1.3±0.2 and a slight y-axis offset of 0.05±0.02 (gray area depicts the 95% confidence band). Only a single measurement exists for Mundrabilla troilite, therefore no error bars can be given for this sample.*

contamination (Palme & Beer, 1993). We have performed different tests to assess whether the Hg concentrations we measured in Orgueil and Mezö-Madaras (another Hg-rich meteorite) could be explained by terrestrial contamination. We have measured fragments of Orgueil both from the surface and the freshly exposed interior of a large (~5 g) fragment, with no significant difference in Hg content. We have pre-heated a powdered sample of Orgueil for 22 h to 100°C on a hot plate, in order to remove possibly adsorbed Hg at low temperatures, but we see no significant difference in Hg content (Table 2). We have also exposed two powdered samples of Orgueil (21 and 25 mg) to the atmosphere for 3 months, at ambient temperature and pressure, to see if the exposed surfaces would trap additional Hg from the atmosphere. The thermal release profile of these two samples (Fig. 1) shows that on the contrary, these two samples have *lost* between 94-97% of their initial Hg. In particular the high-temperature phases at 350°C and >400°C have lost a disproportionally large fraction of their Hg. This suggests that the Hg in Orgueil is primarily contained in a carrier which is normally protected by the surrounding matrix, and is destroyed (or evaporates) upon contact with air. All these observations speak strongly against contamination.

In a second experiment, we have pre-heated a sample of Mezö-Madaras (L3.7) to 150°C, to remove the Hg released at that lowest temperature step in the thermal release profiles (see section 3.2). This lead to the loss of about 50% of the initially present Hg. If Hg released at that temperature step were terrestrial contamination, one would expect it to be terrestrial in isotopic composition, or perhaps enriched in the light (and odd-numbered) Hg isotopes. However, the isotope data for the two Mezö-Madaras samples points in the opposite direction: the Hg from the 150°C step thus has to be enriched in the heavy, and depleted in the odd-numbered Hg isotopes in order to explain the difference between the two isotope measurements (see Fig. 2 and 3). This suggests that Mezö-Madaras might initially have been even more enriched in low-temperature Hg, but some of it was lost to (terrestrial?) volatilization.

Loss of some Hg from a meteorite residing on the Earth's surface can thus not be excluded. All hot desert meteorite finds in our collection (Jbilet Winselwan, CM2; Dhofar 1432, CR2; NWA 7323, LL3; NWA 753, R3.9; and Ghubara, L5) have a low Hg abundance, while the fraction of meteorites with low Hg abundance is smaller for meteorites recovered outside of hot deserts. Meteorites with intact fusion crusts are of dark appearance, which will result in high equilibrium temperatures in desert environments. Loss of other volatile elements (S, Se, Na and occasionally Ni) has been reported for other desert meteorites (Dreibus et al., 1995). Our tentative conclusion is that rapid recovery of meteorite samples is crucial for a reliable Hg analysis, and desert meteorites should not be used.

### 3.2. Thermal release profiles of Hg

All meteorites for which we analyzed thermal release profiles (Fig. 4), without exception, contained a significant low-temperature component released in the lowest temperature step at 150°C (due to technical limitations of the DMA, lower temperature steps are not possible). For most meteorites, this is the largest fractional release step, with 10-80% of total Hg being released. As we will show later, this low-temperature component is slightly depleted in vapor-phase Hg (at least for the L3.7 chondrite Mezö-Madaras), which could suggest that a part of the Hg from this component was already lost to evaporation prior to measurement, either on Earth or in space. Given that Hg has often been associated with sulfides and S release (Lauretta et al., 2001), it seems logical to associate the Hg release in the 150°C step with the phase transition of troilite to pyrrhotite at ca. 100-140°C (Yund and Hall, 1968; Herndon et al., 1975). Trapped Hg in troilite might be released during re-ordering of the crystal structure. Since most of the troilite in the CI1-chondrite Orgueil has already been oxidized to magnetite and pyrrhotite (Herndon et al., 1975), the fractional Hg release should thus be much lower for Orgueil compared to other meteorites, which is indeed the case (Fig. 4). Also, Mundrabilla (IAB) troilite should show a pronounced release of Hg in that step, which is observed as well. On the other hand, Millbillillie (eucrite) shows the most significant release (~80%) at 150°C, but no release at later steps thought to be related to troilite. This might suggest that there are additional carrier phases releasing their Hg at the 150°C temperature step.

The fractional Hg release at the second temperature step, at 200°C, is significantly lower than in the first step for all meteorites: less than 10% of total Hg is released in that step. As discussed in section 3.1., contaminating, adsorbed terrestrial Hg is expected to be released primarily at this step (see "$Hg_c$"-labelled arrow in Fig. 4), so the low fractional abundance at this step suggests that adsorbed terrestrial Hg is a minor (<10%) component in the Hg inventory of the meteorites. This is also supported by a nearly constant proportion between the amounts of Hg released in the first and second steps, respectively, suggesting that the second step is likely a high-temperature tail of the low-temperature component released at 150°C, rather than a separate component with peak release in the 150-200°C interval.

All meteorites show an increased fractional release again at the next temperature step, 250°C, with the possible exception of the Mundrabilla (IAB) troilite sample. This confirms the earlier finding of Lauretta et al. (2001) that a major Hg component in the Allende meteorite is released at ca. 225°C. Our sample of Allende also shows a significant Hg release (~22% of total) in the 200-250°C interval. For most meteorite samples from this study, this is the second-most important release of Hg after the initial step at 150°C. A notable exception is Orgueil, where most of the Hg is released at higher temperatures. The peak at 250°C coincides with major fractional release peaks in the thermal release profiles determined for the synthetic Hg compounds HgS, HgSe and HgO by Lauretta et al. (2001), which are indicated with arrows in Fig. 4. This would suggest that the Hg

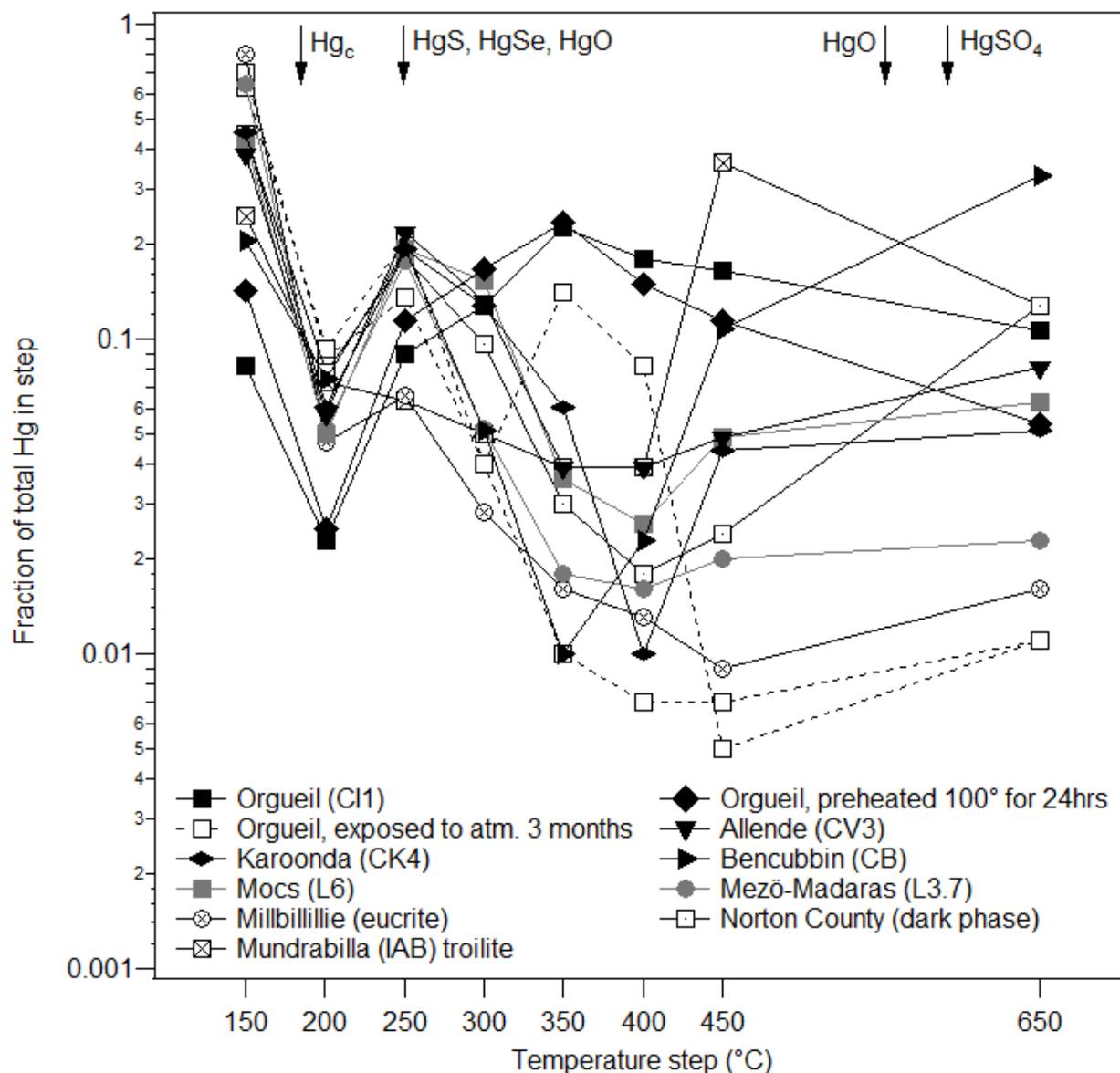

*Figure 4: Release of Hg (as fraction of total Hg in sample) in different temperature steps. At the top of the Figure, peak release positions of different Hg compounds as determined by Lauretta et al. (2001) are given, including "contaminant" terrestrial Hg (Hgc) adsorbed under saturated Hg-atmosphere onto pre-baked Allende to simulate external contamination.*

released in the 250°C step is likely derived from the thermal destruction of Hg compounds.

Most meteorites show a lower fractional release in the next temperature step, 300°C, although for some meteorites this is more pronounced than for others, perhaps suggesting the presence of an additional, unidentified component in Orgueil, Mocs, Mezö-Madaras and perhaps the Norton County dark phase (an oldhamite-rich lithology; e.g., Wheelock et al., 1994). The downward trend continues for the temperature step at 350°C, with the exception of Orgueil, which has its highest fractional Hg release (ca. 24%) in this step. A peak release at 350°C coincides well with the major release peak for Murchison (at 344°C) determined by Lauretta et al. (2001), suggesting that the main Hg carrier phase in Orgueil and Murchison could be the same. Lauretta et al. (2001) suggested a sulfide-related carrier based on the observation of parallel release of Hg and S. In contrast to Lauretta et al. (2001) however, we do not find a similarly pronounced peak for Allende at 350°C. Also, most of the other meteorites seem to be lacking a particular release of Hg at this step, perhaps suggesting that the Hg carrier phases in Orgueil / Murchison on one hand, and all other meteorites on the other, are complementary in the sense that they cannot coexist. Both Orgueil and Murchison are meteorites which have been aqueously altered at low temperatures, while all meteorites lacking the 350°C peak are either pristine or thermally altered. Therefore, one possible explanation is the release of elemental $Hg_{(0)}$, originally produced from the oxidation of Hg-bearing sulfides during aqueous alteration. Elemental Hg, which has a boiling point of 357°C at ambient pressure, has also been found in the H3.6 chondrite Tieschitz (Caillet Komorowski et al., 2012), where it is found in association with Fe- and Cu-sulfides, native Cu and cinnabar. A third candidate for Hg release in this step is the transition between two allotropes of HgS, from cinnabar ($\alpha$HgS; trigonal-trapezohedral) to metacinnabar ($\beta$HgS; isometric-hextetrahedral) at approximately 345°C (Sharma & Chang, 1993).

The 400°C step brings another reduction of fractional releases for all meteorites (including Orgueil). For Karoonda (CK4), the drop in fractional release is dramatically higher for this step than for other meteorites, perhaps suggesting that there is a specific Hg carrier in this interval which is present in most meteorites, but not in Karoonda. A separate carrier in this interval is also supported by the release profile of Bencubbin (CB), which shows a higher fractional release in this step. The 450°C step is probably the most interesting since different meteorite samples show clearly different behavior. The highest fractional release is by Mundrabilla troilite, which releases almost 50% of its total Hg inventory in this step. This suggests that the main carrier releasing Hg at this step is troilite. Apart from the Mundrabilla troilite, also Bencubbin and Karoonda both show a strong increase of fractional release of Hg, while Mocs (L5-6), Mezö-Madaras (L3.7), Allende (CV) and the Norton County dark phase show a slight increase. The fractional release of Hg from Orgueil further decreases in this step. In the final step at 650°C, three meteorites release significant fractions of their total Hg: Bencubbin (33%), Mundrabilla troilite and Norton County (both 13%). All other samples release the final <10% of their Hg in this step. Comparing with the thermal release profiles of synthetic materials by Lauretta et al. (2001) again, possible carrier phases at this step include HgO and $HgSO_4$. In Table 3, we summarize all meteorites, temperature steps and candidate carrier phases.

The 650°C step is the highest one achievable with the DMA. The question remains whether more Hg-carrying phases would release their Hg at even higher temperatures, in particular organic compounds, given the organophilic nature of Hg. The likely carbon-rich "phase Q" noble gas carrier releases its trapped noble gas contents only at surprisingly large temperatures (1000-1600 °C; Huss et al., 1996), so it is at least conceivable that Hg, which is less volatile than noble gases, could still be released at very high temperatures too. Given that the variation between Hg concentrations measured by DMA and HPA extraction is not larger than the typical between-sample variability (~10-20%), the contribution of a high-temperature (possibly organic) carrier phase to the total Hg content is unlikely to be much higher than this. An unknown, additional Hg carrier in meteorites that is both hot acid- (HPA) and temperature (DMA) resistant, and which would thus have escaped our analysis, can logically not be excluded, but is also not called for.

### 3.3. Hg abundance variation

The Hg abundance of a meteorite is not a simple function of its thermal history (Fig. 1). Some meteorites which should have experienced only little parent body heating/processing (e.g., the CR2 chondrite Dhofar 1432 or the LL3 chondrite NWA 7323) have very low Hg abundances on the order of 10 ppb, less than the bulk Hg abundance of some achondrites which have seen thorough thermal processing, e.g. the ureilite Almahata Sitta (200 ppb) or the aubrite Norton County (90 ppb). As mentioned in section 3.1., in the case of desert meteorites, it is likely that low abundances are due to terrestrial Hg loss. But in the case of Tagish Lake, which was recovered from a frozen Canadian lake within no more than a few months of the fall, yet still has a concentration of only 70 ppb Hg, this is an unlikely explanation. The two highest Hg concentrations measured in this study are from the CI1 chondrite Orgueil (ca. 14'000 ppb) and the CM2.7 chondrite Paris (ca. 11'000 ppb). Hg thus seems to be one of the very few elements which is enriched in Orgueil relative to its abundance in the solar photosphere. While Orgueil is a highly aqueously altered meteorite, Paris is considered to be almost "pristine" (i.e., thermally and aqueously unaltered; Hewins et al., 2014; Marrocchi et al., 2014). Therefore, the Hg concentration is also not a function of aqueous alteration either. If anything, only the upper envelope of the Hg abundance variation is decreasing towards higher petrographic types – but even this trend hinges on the high abundances we found for Orgueil and Paris. Wiederhold & Schönbächler (2015) report a Hg abundance of >10'000 ppb in a sample of Allegan (H5), demonstrating that surprisingly high Hg abundances are also found in petrographic types >3. While maximum to minimum ratios of ~1000 have been observed for other moderately to highly volatile elements in meteorites, the abundances of these elements do generally correlate with petrographic type (e.g., Wasson & Kallemeyn, 1988; Wasson, 1972), which is not the case for Hg.

Mercury concentrations in meteorites do not show a particular preference for the Hg bulk solar system silicate ($Hg_{BS3}$) value of approximately 260 ppb suggested by Lauretta et al. (1999). Only two meteorites from our suite, Kainsaz (Ornans-type; CO3.2) and Bencubbin (Bencubbinite; CB) have a Hg concentration that falls within error of $Hg_{BS3}$, while the values of Karoonda (Karoonda-type; CK4) and Almahata Sitta fall within a factor of two of that value. The $Hg_{BS3}$ value is also relatively close to the Hg concentration found for Murchison (CM2) by Lauretta et al. (2001). Our sample of Murchison, however, had a significantly higher Hg concentration of ~2500 ppb. Variable Hg concentrations in different samples of the same meteorite have been observed previously. For Orgueil, concentrations of between 500 and 500'000 ppb have been reported from chemical analysis and NAA studies (see Lauretta et al., 1999, and references therein). Some of the variation observed in NAA studies can likely be explained with interference from [75]Sn and [203]Hg (Ebihara et al., 1998; Kumar et al., 2001), but even mass spectrometry-based studies like Lauretta et al. (2001), Wiederhold & Schönbächler (2015) and this work confirm that a range of different Hg concentrations exists, even within different samples of the same meteorite (a factor of ~10 for Murchison and Orgueil, and a factor ~3 for Allende).

If terrestrial contamination can be excluded, as we conclude from our series of experiments (see section 3.1.), the observation of variable Hg concentration between samples of the same meteorite requires that Hg shows a mobile behavior on different meteorite parent bodies, and can be locally enriched by orders of magnitude, at least on the level of centimeter- to meter-sized samples. Lauretta et al. (2001) demonstrated that HgS is likely a main carrier of Hg in Murchison. Palme et al. (1985) reported an unusual, sulfide-rich fragment found in a cut of the Allende meteorite, which was enriched in Hg compared to bulk Allende. Caillet Komorovski et al. (2012) report the discovery of heterogeneously distributed HgS and Hg in the matrix of the Tieschitz (H3) meteorite. Although Caillet Komorovski et al. (2012) report Hg abundances of up to 25'000-30'000 ppb for their sample of Tieschitz, a small (4.7 mg) sample of the same sample (provided by Ahmed El Goresi) analyzed with the DMA in Nancy yielded only ~30 ppb Hg (Fig. 1, Table 1). Strongly heterogenous distribution of Hg-carrying phases can thus explain the variable concentrations of Hg found in different fragments of the same meteorites. This makes it challenging to estimate a true bulk concentration of Hg, but a series of independent measurements in different fragments all suggesting abundances significantly above the $Hg_{BS3}$ value (as is the case, e.g., with Orgueil) have to be interpreted as a bulk enrichment of Hg within certain meteorites.

### 3.4. Mass-dependent fractionation of Hg isotopes in meteorites

While we find large variations of up to ~7‰ in $\delta^{202}Hg$ (ca. -1.8 ‰/amu), this variation does not correlate with Hg abundance (Fig. 2). Co-variation would be expected if the strong variation in Hg concentrations of meteorites were the consequence of a single fractionation process. For example, kinetic fractionation (evaporation of Hg into vacuum) would induce a strong fractionation of mass-dependent Hg isotope ratios even if only a small fraction of total Hg is lost (as shown by the nearly-horizontal arrows radiating from the data points of Orgueil and Kainsaz in Fig. 2). Evaporation under closed-system, equilibrium conditions induces less fractionation for a given Hg loss (as shown by the arrows with steeper slopes in Fig. 2). These relationships can be used to test scenarios for the origin of Hg abundances and isotopic compositions in meteorites: The arrows in Fig. 2 show that it is not possible, e.g., to reach both the Hg concentration and isotopic composition of Orgueil by starting at $\delta^{202}Hg = 0$ at the $Hg_{BS3}$ concentration (260 ppb), by kinetic fractionation alone. Such a scenario is only possible if equilibrium fractionation is dominant. However, mass-dependent fractionation under dynamic conditions is also strongly temperature-dependent (Estrade et al., 2009), and will thus be zero (i.e., the fractionation line will be vertical in Fig. 2) for temperatures >115°C. Interestingly, the two meteorites with a Hg abundance closest to the $Hg_{BS3}$ value, Kainsaz and Bencubbin, can be connected by a kinetic fractionation line, perhaps suggesting that some meteorites which did not experience major relocation of Hg on their parent bodies had their $\delta^{202}Hg$ values initially set by kinetic fractionation, which could be a signature of nebular condensation. The Hg abundance and mass-dependent Hg isotope fractionation of a meteorite might thus be established in at least two steps: (1) capture and kinetic fractionation of Hg in the early solar system, followed by (2) enrichment / depletion of Hg driven by asteroidal thermal evolution.

### 3.5. Mass-independent fractionation of Hg isotopes in meteorites

Mass-independent fractionation of Hg isotopes has been observed in a variety of terrestrial environments, for both odd-numbered (199, 201) and even-numbered (200, 204) isotopes (see Blum et al., 2014, for a review). In our suite of meteorites, we find no significant mass-independent variation for the even-numbered Hg isotopes (internal renormalizations to 200, 202, 204, or to an average of all three even-numbered isotopes yield the same mass-dependent isotopic fractionation, within uncertainty). However, we find pronounced mass-independent fractionation in the odd-numbered isotopes, of up to ~0.3 permil (Table 1). The overall slope of the correlation line between $\Delta^{199}$Hg and $\Delta^{201}$Hg in Fig. 3 (1.30±0.19, 1$\sigma$) is compatible with both kinetic fractionation (evaporation against vacuum in an open system where the saturation pressure of Hg is never reached; slope 1.2±0.4; Estrade et al., 2009) and equilibrium fractionation (evaporation in a closed volume; slope 2.0±0.6, Estrade et al., 2009; 1.59±0.05, Ghosh et al., 2013), with kinetic fractionation slightly favored. As shown in Fig. 3, nearly all data points with mass-independent isotopic enrichments are from carbonaceous chondrites (the exception being a single data point from Mundrabilla troilite). On the other hand, ordinary chondrites and achondrites show values similar to the Earth, or – in the case of the Millbillillie eucrite, even show depletions in the odd-numbered Hg isotopes. Only a single carbonaceous chondrite (the CK chondrite Karoonda) shows a depletion in odd-numbered isotopes. Interestingly, the CK chondrites are among the few carbonaceous chondrite groups having experienced significant thermal alteration. Therefore, a connection between MIF effects and thermal/aqueous alteration seems likely.

## 4. Discussion

### 4.1. Absence of nucleosynthetic anomalies on $^{196}$Hg

Correlated variation of $\Delta^{199}$Hg and $\Delta^{201}$Hg values (as discussed in the next section) can be due to either MIFs or the presence of varying fractions of nucleosynthetic Hg (based on the supernova models from Rauscher et al., 2002). Since SN-derived Hg should also show correlated enrichments of $^{196}$Hg on the order of several 1000‰, $^{196}$Hg can be used to test the hypothesis that the variation in odd-numbered isotopic ratios is due to the admixture of supernova material. For example, as shown in Fig. 5, a $\Delta^{199}$Hg = +0.1 should result in a $\Delta^{196}$Hg = -2. This is however not observed: all meteorites for which $^{196}$Hg was measured plot within uncertainty of the terrestrial $\Delta^{196}$Hg value.

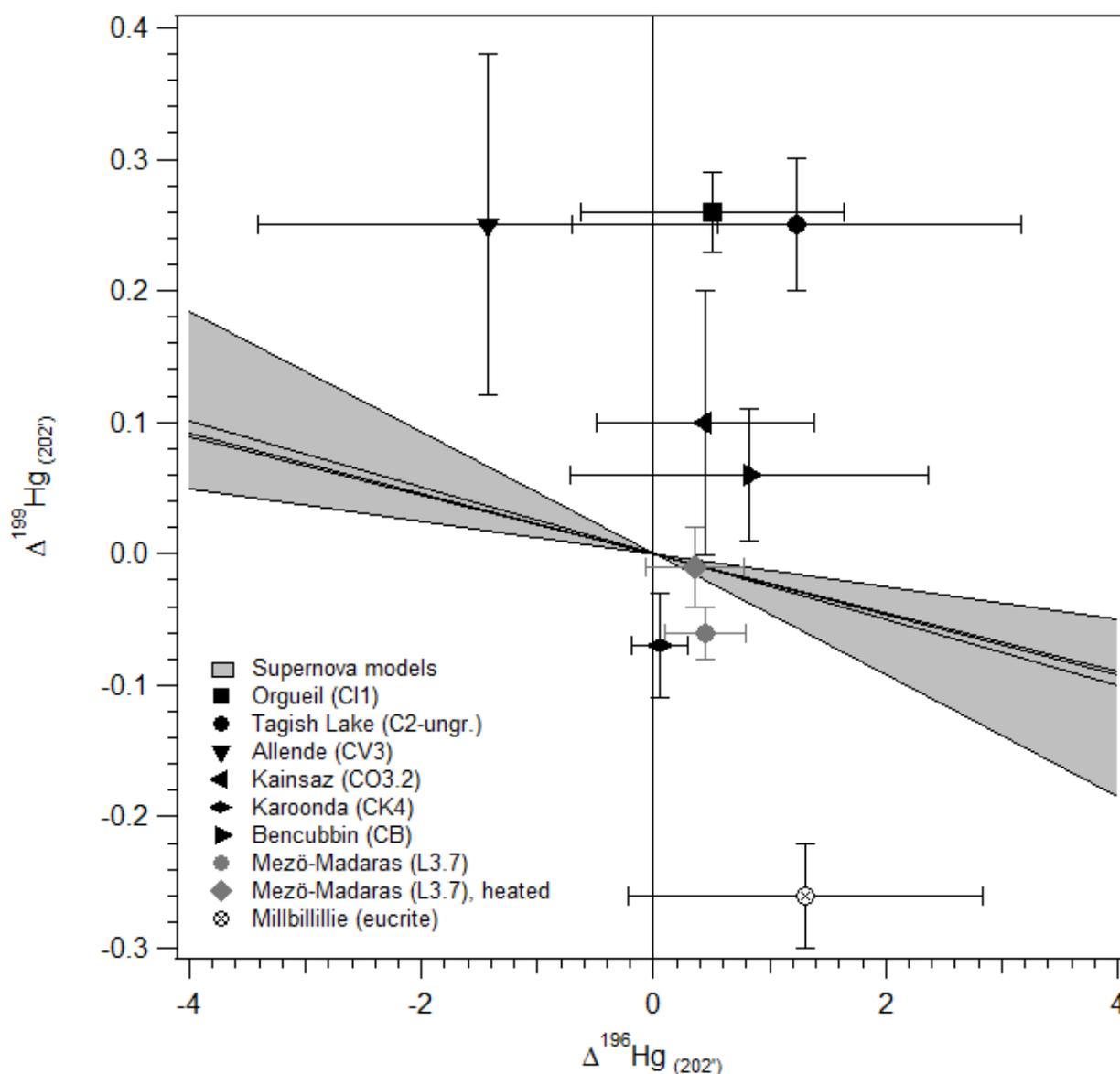

*Figure 5: If the observed Δ199Hg variation were of a nucleosynthetic (supernova) origin, we would expect co-variation between Δ199Hg and Δ196Hg (based on the SN-models by Rauscher et al., 2002), i.e. the data points should fall within the gray shaded area, but this is not observed.*

Furthermore, no significant variation in $\Delta^{196}$Hg is seen between the pristine and the pre-heated sample of Mezö-Madaras (L3.7), suggesting that none of the two Hg components (the one released at 150°C, and the other at >150°C, respectively) contains a significant contribution of nucleosynthetic Hg. The error-weighted average $\Delta^{196}$Hg of all meteorites is 0.24±0.34, i.e. the Earth and meteorites have an identical $\Delta^{196}$Hg value within uncertainty. While small contributions of nucleosynthetic $^{196}$Hg to each of the meteorites cannot be excluded at this point, they are clearly not the dominant source of variation in the odd-numbered isotope abundances. The solar system was thus well-mixed with respect to late additions of supernova-derived Hg potentially added at a late stage. This is in agreement with observations of other moderately to highly volatile elements, showing no isotopic variation in different solar system materials (Yokoyama et al., 2014).

### 4.2. Evaporation and re-condensation of Hg in asteroids heated by primordial radionuclides

In any asteroid reaching temperatures above the melting point of Hg (and other Hg-carrying compounds) by heating induced by the decay of primordial radionuclides (mostly $^{26}$Al, $t_{1/2} = 0.7$ Ma), Hg will start to evaporate into internal pore volumes. Then, some Hg migration along pressure- and temperature-gradients – generally from the hot interior towards the cold crust – must occur (as previously suggested for other volatile elements, e.g., Cd, by Wombacher et al. 2003; 2008). During this process of continuous evaporation and re-condensation driven by the thermal evolution of the asteroid, it is conceivable that this will lead to isotopic fractionation because Hg is only partially evaporated, or because vapor-phase Hg (enriched in light and odd-numbered isotopes) is preferentially condensed in the cold crust of the asteroid. Additional isotopic fractionation processes, e.g., photochemical reduction of near-surface Hg by irradiation by the sun, oxidation/reduction reactions of Hg with other elements, likely S (c.f. Schaefer & Fegley, 2010; Smith et al., 2014), or incorporation of Hg into organic compounds might also be at work. For our first-order model, we considered only evaporation and re-condensation of elemental Hg (Hg$_{(0)}$), and leave more refined models to future work. We based our model on the asteroid heating model originally developed by Miyamoto et al. (1981) and later updated by Bennet & McSween (1996). We implemented both an "ordinary chondrite" and a "carbonaceous chondrite" model using different asteroid radii, initial temperatures, densities, porosities and formation times. For the carbonaceous chondrite model, we performed a parameter study in order to search for the conditions leading to maximal isotopic fractionation (see below). The heating model is described in detail in the Supplementary Online Materials, while the results are shown in Fig. 6, 7, and 8. The left axis of Fig. 6 and 7 doubles as a (logarithmic) scale of maximum temperature (K) and Hg concentration (ppb): The gray shaded area shows the final concentration of Hg at the end of thermal evolution (when the asteroid interior reaches ambient temperature, ca. 500 Ma after formation of the first condensates in the solar system, Calcium-Aluminum-rich inclusions or CAIs, usually taken as "time zero" for early solar system studies), the dashed orange curve depicts the maximum local temperature ($T_{max}$) reached. The $\delta^{202}$Hg and $\Delta^{199}$Hg curves (solid red with white and red markers, respectively) are given in permil and refer to the right axis. All values are plotted versus the depth below the surface of the asteroid on the lower axis (asteroid center to the left, near-surface to the right of the logarithmic scale).

The interior of the (R = 85 km, $T_{init}$ = 190 K, $\rho$ = 3.6 g/cm$^3$, porosity = 2.5%, $T_0$ = 2.1 Ma after CAI; following Sugiura & Fujiya, 2014) ordinary chondrite parent asteroid is devoid of Hg at the end of thermal evolution. At about 1.2 km depth (corresponding to a maximum temperature of ~370 K), a strong increase in Hg abundance takes place, climbing to almost 15'000 ppb near the surface (Fig. 6), which is more than measured in any ordinary chondrite from this work. Smaller asteroid radii lead to lower surface concentrations, but show the same general picture. While the model assumes no Hg loss to space, continual escape of, e.g., 50% of the Hg redistributed into the uppermost shell (in direct contact with the surface) does not lead to significantly lower total Hg (see dashed line in Fig. 6). The cold (below the solidus of elemental Hg) surface acts as a "cold trap", blocking Hg loss into space. It is possible that heating of the surface by impacts would change that (Ciesla et al., 2013), but this is difficult to constrain and has thus not been considered here. Due to the almost complete mobilization of asteroidal Hg in these models, mass-dependent and mass-independent isotopic fractionation is limited (within ~0.02‰ for $\Delta^{199}$Hg, and within ~0.25‰ for $\delta^{202}$Hg; note that these values are relative to the initial, which does not necessarily have to be identical with the terrestrial standard). The Hg isotopic composition of Hg-rich ordinary chondrites might thus be very close to their initial Hg isotopic composition. The strongest isotopic fractionation effect (a strong enrichment in heavy isotopes) is observed at the base of the Hg-rich layer, where Hg concentrations are comparatively low. The models are successful in reproducing the high (>Hg$_{BS3}$) Hg enrichment we measured in the L3.7 chondrite Mezö-Madaras (3150±520 ppb) and the L5-6 chondrite breccia Mocs (965±140 ppb). However, given their petrographic types, both meteorites must have seen higher temperatures during their thermal metamorphism than the <370 K found in the Hg-rich, outer-most crust (~900 K for L3.7 chondrites, ~1300 K for L5-6 chondrites; Wlotzka, 2005; Bennet & McSween, 1996). This can be explained if the ordinary chondrite parent body was disrupted and re-accreted during the time of Hg mobilization. Pre-heated samples would then be re-deposited close to the surface, where they are subsequently enriched in Hg after thermally induced migration continues. Impacts on the surface will also lead to regional heating, which could induce re-mobiliziation of Hg long after the thermal evolution of the asteroid has ended, but as long as the resulting temperature stays below 115°C, this would also induce Hg isotopic fractionation (MIF and MDF), which is not observed for ordinary chondrites.

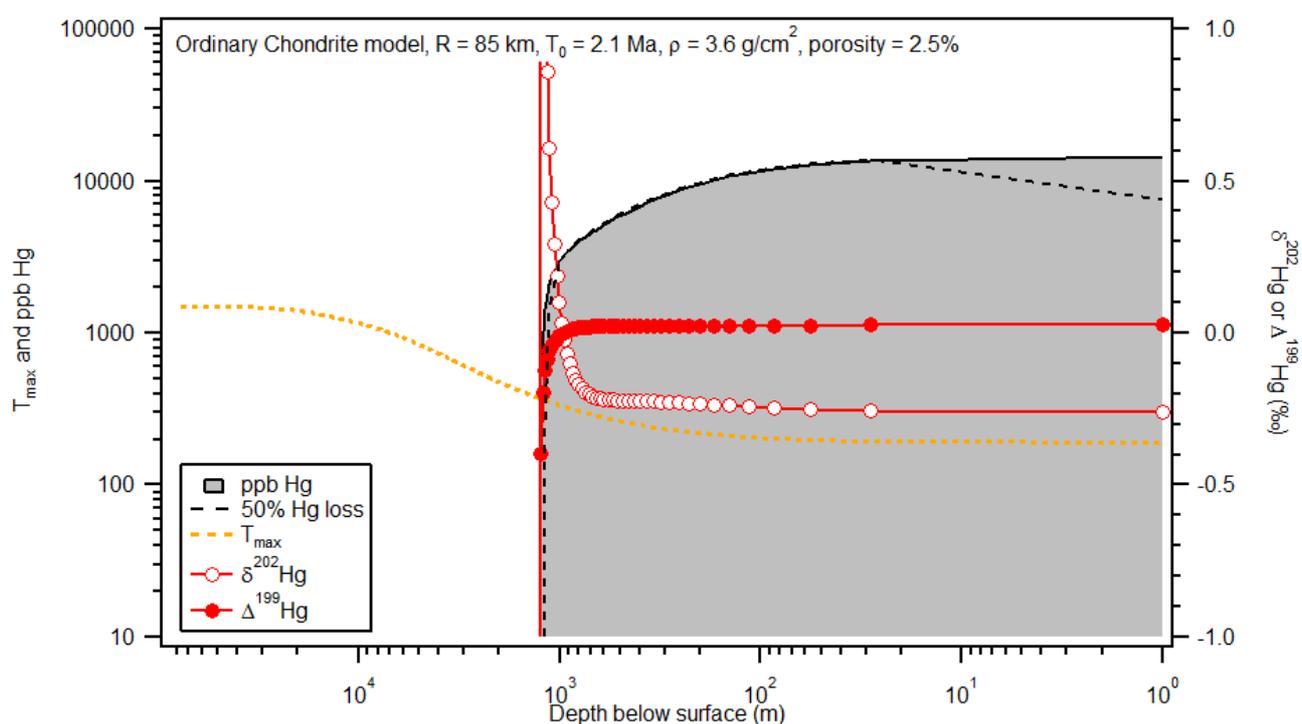

*Figure 6: The shaded area shows the distribution of Hg at the end of thermal evolution, as a function of depth within the asteroid (center to the left, surface to the right). Here, the asteroid is of ordinary chondritic composition, the porosity is 2.5%, the radius is 85 km, the initial and the ambient temperature are 190 K, and the formation time is 2.1 Ma after CAIs. Also shown are the concentration values measured in some ordinary chondrites. The left axis doubles as temperature axis, the maximum temperature reached at a given depth (Tmax) is given as an orange dashed line. The red curves with white and red markers (δ202Hg, Δ199Hg, respectively) give the isotopic composition of the Hg deposited at a given depth at the end of thermal evolution relative to initial (δ202Hg = Δ199Hg = 0).*

In order to reproduce the stronger isotopic variation (in particular, the MIF in $\Delta^{199}$Hg) seen in many carbonaceous chondrites, we conducted a parameter study, varying either the formation age (3, 3.5, and 4 Ma after CAIs, following again Sugiura & Fujiya, 2014), porosity (5, 10, 20, 50%) and radius (10, 20, 40, 80, and 160 km) of the parent body, while keeping the other parameters constant. This shows that the maximum isotopic variation (at the surface) expected from the model is on the order of 1.2‰ (for $\delta^{202}$Hg) and 0.12‰ (for $\Delta^{199}$Hg), while the Hg concentration at the surface varies from 260 ppb (no significant change relative to initial, for late-formed small bodies) to 25'000 ppb. Larger asteroids have higher near-surface concentrations, while later formation and higher porosity lead to higher isotopic fractionation, with a peak in fractionation observed for asteroids in the 10-40 km range (see Fig. 8). In Fig. 7, we show the results for the carbonaceous chondritic asteroid yielding the strongest isotopic fractionation (R = 20 km, $\rho$ = 1.5 g/cm$^3$, porosity = 50%, T$_0$ = 4 Ma after CAI). Such high porosities (including both micro- and macro-porosity) are possible if the parent body is a rubble pile similar to the C-type asteroid (253) Mathilde, which has a mean radius of 26 km and a bulk density of 1.3±0.2 g/cm$^3$ (Veverka et al., 1997). A radius of 20 km is also the smallest for which the orbit of the asteroid is not significantly affected by non-gravitational forces over the history of the solar system (Bottke et al., 2006), i.e., asteroids with ~20 km radius are the smallest possible "primordial" asteroids still surviving today, even if smaller primordial asteroids may have existed initially (e.g., Weidenschilling, 2011). Meteorites falling on Earth today must therefore be derived from primordial asteroids (or their fragmented remains) >20 km in radius. Apart from the stronger isotopic fractionation, the general picture is similar to the ordinary chondritic model: the innermost ~14 km of the asteroid are devoid of Hg, and the strongest isotopic fractionation is seen at the base of the Hg-rich, surface-near layer (however, the Hg-rich layer is thicker at 6 km in this model compared to 1.2 km in the 85 km ordinary chondritic model). The maximum temperature reached at the depth where the Hg abundance increases rapidly is 290 K. Again, this is within the range of temperatures expected for meteorites having experienced only mild thermal alteration (e.g., Kainsaz), while other meteorites must have seen higher temperatures (e.g., Karoonda), suggesting again sample mobilization during thermal metamorphism, or local Hg re-mobilization at a later stage.

In Fig. 8, we plot the results of our parameter search study together with the measured data ($\Delta^{199}$Hg and concentration in ppb) in meteorites. By plotting all carbonaceous chondrites into the same diagram, we assume that their initial $\Delta^{199}$Hg was similar, as suggested by the absence of nucleosynthetic anomalies in Hg and other moderately volatile elements (see section 4.1.). Four of the meteorites (Bencubbin, Kainsaz, Karoonda, Murchison) have values which are compatible with the model results within uncertainty, i.e., their Hg concentration and mass-independent isotope fractionation can be understood to be caused by Hg evaporation and re-condensation on the (10-40 km radius) parent asteroid. Note however that some of these meteorites show significant differences in mass-dependent isotope fractionations, which might have been set already in the solar nebula (see section 3.3.). The high Hg concentrations measured in Orgueil and Paris are reached in the surface-near layers of some asteroids with large radii. According to the model, this surface-near Hg should not be isotopically fractionated in mass-independent isotopes relative to the initial, but it clearly is in Orgueil and Paris (Fig. 8). Therefore, at least one additional mass-fractionating process is required, e.g., the addition of more vapor-phase Hg beyond what can be provided in the asteroid heating model. Typically, vapor-phase Hg has a $\Delta^{199}$Hg value of ~0.12 relative to the Hg from which it was derived (Estrade et al., 2009), i.e., to attain the isotopic composition of Orgueil and Paris (+0.26±0.03 and 0.31±0.06, respectively), 2-3 enrichment steps are required. Tagish Lake and Allende, which have low Hg abundances, but also show an excess in $\Delta^{199}$Hg not explicable with the heating model, may have experienced partial Hg loss after initial enrichment in vapor-phase Hg. As already noted by Caillet Komorovski et al. (2012), parent body heating alone is unlikely to lead to multiple pulses of Hg loss and redeposition. Such a behavior could instead be expected in the context of repeated impact heating.

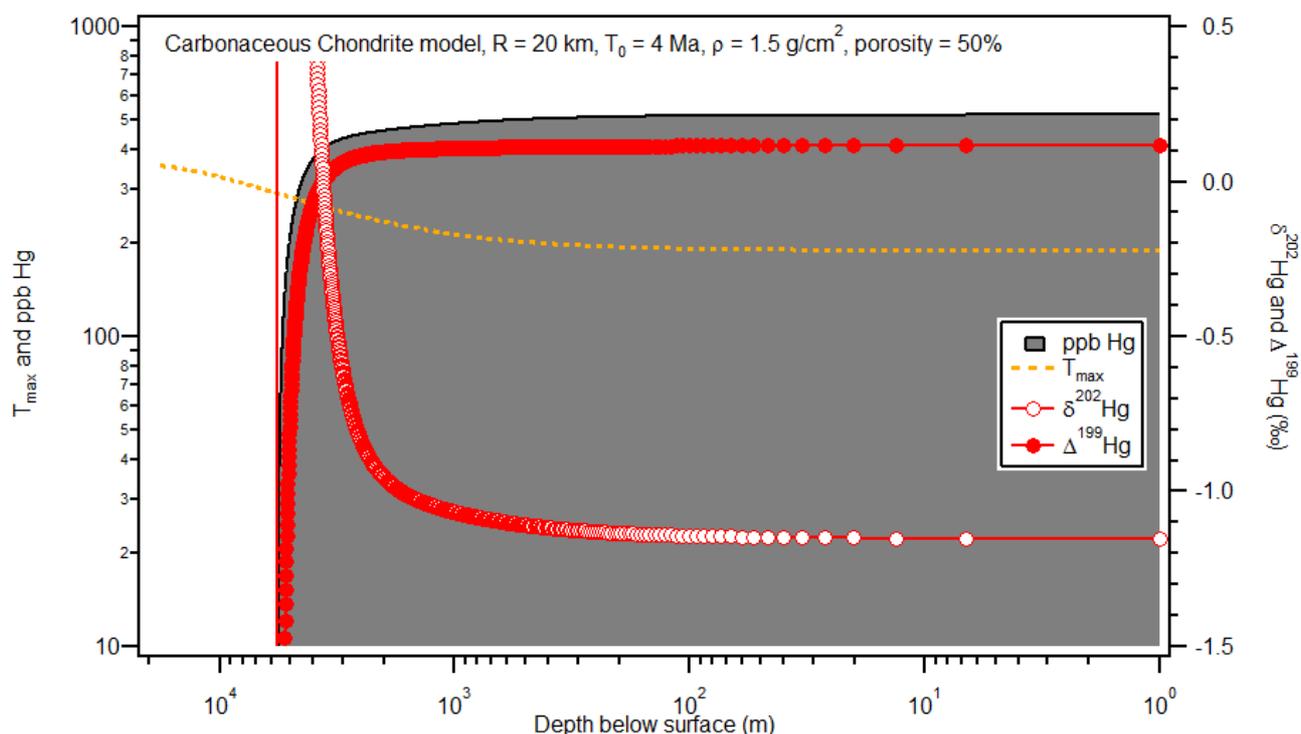

*Figure 7: Same as for Figure 6, but here, the asteroid is of carbonaceous chondritic composition, the porosity is 50% (i.e., the density is 1.5 g/cm3), the radius is 20 km, the initial and the ambient temperature are 170 K, and the formation time is 4 Ma after CAIs.*

### 4.3. Hg in icy deposits – and comets

A cometary origin has previously been suggested for Orgueil (Gounelle et al., 2006) and for some CM-chondrites (Seargent, 1990; Haack et al., 2011; Meier, 2014). It is thus interesting to note that both Orgueil and the primitive CM chondrite Paris show high enrichments in both total Hg and vapor-phase Hg. Another possible clue for a connection between ice-rich environments and Hg comes from icy deposits at the lunar poles. High Hg abundances in these deposits were suggested by Reed (1999), and confirmed in the plume ejected from the floor of the lunar polar crater Cabeus by a kinetic impactor (the LCROSS mission; Gladstone et al. 2010). The lunar soil in this crater was found to contain 2'000'000 ppb Hg! The lunar polar ice deposits are thought to form by the evaporation of volatile species on the surface of the Moon, followed by migration into these cold traps on the permanently shadowed crater floors in the lunar polar regions. The Hg in these deposits is likely also enriched in vapor-phase Hg, with a $\Delta^{199}$Hg vs. $\Delta^{201}$Hg slope determined by kinetic fractionation. A process of evaporation and redeposition, similar to the one on the Moon, might have taken place on the parent bodies of Orgueil and Paris. If these bodies were indeed comets, we would certainly expect the meteorites to have gone through multiple cycles of surface volatile evaporation and re-condensation. Since Hg has a high mass and a relatively high first ionization potential, its is more likely to be retained by the comet relative to other volatiles (e.g., water). In addition, it has been suggested that comets contain a thin surface layer enriched in organic matter (Capaccioni et al., 2015), providing yet another site where the organophilic Hg might be trapped and enriched. Unfortunately, direct determination of Hg in the coma of comet 67P/Churyumov-Gerasimenko by the Rosetta space probe is not possible. Clearly, no firm conclusion can be drawn from these observations at this time, but more work on the sites of Hg storage in Orgueil and other Hg-rich meteorites (and eventually, lunar polar soils) is needed.

### 4.4. The origin of Hg on Earth

How does the Earth's Hg isotopic composition compare with the rest of the solar system? Terrestrial Hg has a mean $\delta^{202}$Hg of -0.68±0.45‰ (Blum et al., 2014), which is heavier than all meteorites analyzed in this study. On the other hand, terrestrial Hg also takes an intermediate position within the trend of mass-independent isotope fractionation (Fig. 3), being more depleted in odd-numbered Hg isotopes than carbonaceous chondrites, but similarly or more enriched in odd-numbered Hg isotopes than ordinary chondrites and most achondrites. This can also be seen in a plot of mass-independent vs. mass-dependent fractionation (Fig. 9). If evaporation and re-condensation were the only processes responsible for the isotopic fractionation of Hg in meteorites, and if the initial isotopic composition of Hg were the same for all meteorites (the latter being suggested by the thorough mixing implied by the absence of nucleosynthetic anomalies, discussed in section 4.1), all meteorites should plot on a single line with a slope close to ~-0.1 in that diagram. This is not the case, although meteorites with very light Hg isotopic composition are preferentially enriched in the odd-numbered isotopes, and vice versa. The three L-chondrite samples (Mocs, Mezö-Madaras and Mezö-Madaras devolatilized at 150°C) clearly plot on a single line with a slope of ~-0.1, as do the two samples from the Norton County breccia. Projecting the measured $\delta^{202}$Hg values of all meteorites along a -0.1 slope onto $\Delta^{199}$Hg = 0 leads to more than half of the meteorite samples plotting at a similar value (~-1.5‰), with most of the rest plotting at another, isotopically lighter value (~-3.5‰), as shown in the histogram inset in Fig. 9. The same two groups emerge if the measured $\delta^{202}$Hg values are used, although they then consist of different meteorites (e.g., Orgueil has a measured $\delta^{202}$Hg of ~-3.5‰ but a projected $\delta^{202}$Hg of ~-1.5‰). In both cases, the Earth's crustal Hg isotopic composition is heavier in $\delta^{202}$Hg than in most meteorites. Only a few meteorites (the two CM chondrites Paris and Murchison) can be projected directly to the terrestrial $\delta^{202}$Hg value.

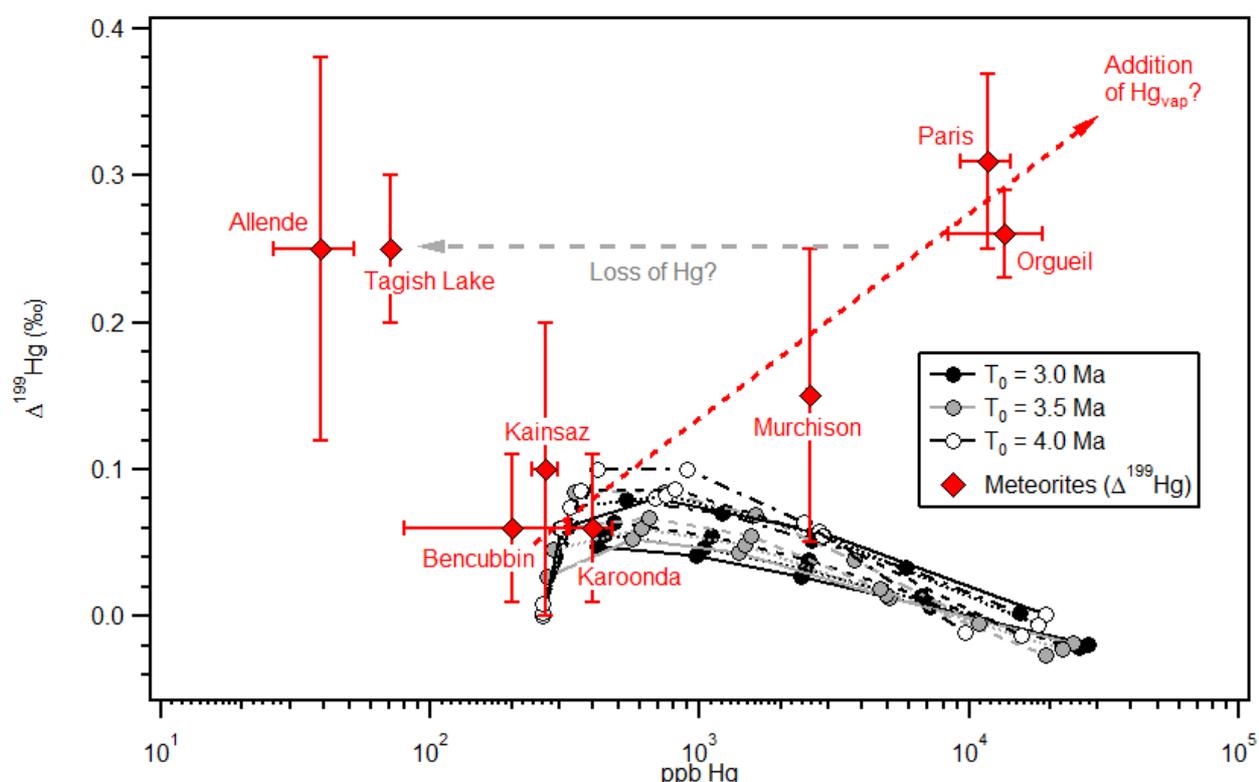

*Figure 8: MIF and Hg abundance in the top-most layer of the asteroid, after 500 Ma of thermal evolution. Solid black dots represent models with a formation age of 3 Ma, grey dots an age of 3.5 Ma, and open dots an age of 4 Ma. Asteroid sizes are proportional to the final Hg abundance, with radii of 10, 20, 40, 80, and 160 km from left to right. Solid lines connect models with a porosity of 5%, while dotted, dashed and dot-dashed-lines connect models with a porosity of 10%, 20%, and 50%, respectively. All carbonaceous chondrites for which the Hg abundance and isotopic composition was measured are given for comparison.*

There are essentially two possible sources for the origin of the Earth's Hg: either, it was retained from the time of accretion ("primordial" Hg from building blocks), or the primordial Hg was lost during the Giant Impact era and the terrestrial Hg inventory was later replenished by late accretion. A third possibility is that both sources contribute a part of Hg on Earth. Primordial Hg is difficult to test for with Hg isotopes, because it is unclear how strongly the suspected loss of Hg during initial accretion and Giant Impacts would affect isotopic fractionation. Values for residual primordial terrestrial Hg of up to +50‰ ($\delta^{202}$Hg) relative to initial Hg (at 260 ppb) are possible with purely kinetic fractionation and a present-day bulk Earth Hg abundance of 10 ppb (within a factor of 4; MacDonough & Sun, 1995; see also Canil et al., 2015), and even more if only a fraction of the present-day Hg is primordial. This is far in excess of the values seen in meteorites, which can thus not put very strong constraints on the fate of the Earth's primordial Hg.

Can we instead go the other way and try to constrain the possible contribution of "late accretion" to the Earth's Hg budget? Late accretion is discussed for siderophile elements in the Earth's mantle (then called the "late veneer"; e.g., Fischer-Gödde & Becker, 2012), and for other volatile elements and compounds (e.g., water, N, C, noble gases; Marty, 2012). A typical estimate for the mass added to the Earth by the "late veneer" is ~0.35% of an Earth's mass. Ten ppb Hg (within a factor of 4) requires the material of the late veneer to contain 700 – 10'000 ppb Hg (average: 3000), which is higher than measured in most meteorites (and higher than the $Hg_{BS3}$ value) but still in the range observed for some carbonaceous chondrites, like Murchison, Orgueil and Paris. As the bulk Hg concentration of the carbonaceous chondritic parent bodies must be lower, such a scenario is only possible if the late veneer is derived from debris ejected from the Hg-enriched surfaces of parent bodies. This scenario would however also require that the isotopic composition of terrestrial Hg resembles the one found in similarly enriched carbonaceous chondrites, i.e., it should show the same MIF as carbonaceous chondrites. As shown in Fig. 3, this is not the case. The "late veneer" can thus be excluded as a major source of terrestrial Hg.

On the other hand, Marty (2012) suggested that the Earth accreted about ~2±1% of its mass as volatile-rich, carbonaceous chondritic material during late accretion to explain the volatile inventory of the Earth ("late accretion"). That mass fraction is larger than postulated for the late veneer, but might be explained by the accretion of a few "wet" (volatile-rich) planetesimals towards the end of accretion. In that case, terrestrial Hg would sample the bulk Hg of the planetesimal, not only its outer-most layers, and would thus be representative of the initial Hg isotopic composition within the planetesimal, but not of present-day carbonaceous chondrites derived from surfaces enriched in vapor-phase Hg. If the planetesimals contained Hg initially at the $Hg_{BS3}$ abundance, the resulting bulk Hg abundance of the Earth is 3 – 8 ppb, well within the range given by McDonough & Sun (1995) and the values extrapolated for the mantle by Canil et al. (2015). Since surface-near reservoirs of Hg (soils, ocean and atmosphere) only contain about ~6000 Mmol Hg (Mason & Sheu, 2002), the vast majority (~1 – 10$^{-6}$) of terrestrial Hg must still be contained within the crustal rocks and the mantle. If Hg, like S, Se and Te, becomes more siderophile at high pressures (e.g., Rose-Weston et al., 2009), a significant portion might also have been incorporated into the core. This opens the possibility of significant isotopic fractionation between the surface and deep reservoirs. Indeed, Zambardi et al. (2009) found a δ$^{202}$Hg value of -1.74‰ in volcanic emissions from Vulcano island, suggesting that Hg in the mantle, like meteorites, is isotopically light compared to the Hg at the surface. Late accretion, contributing approximately 2% of the Earth's mass (Marty, 2012), can thus explain most of the Earth's Hg, with negligible or minor primordial Hg. The dominant source of Hg on Earth is therefore likely to be the same as for water, nitrogen, carbon and the noble gases.

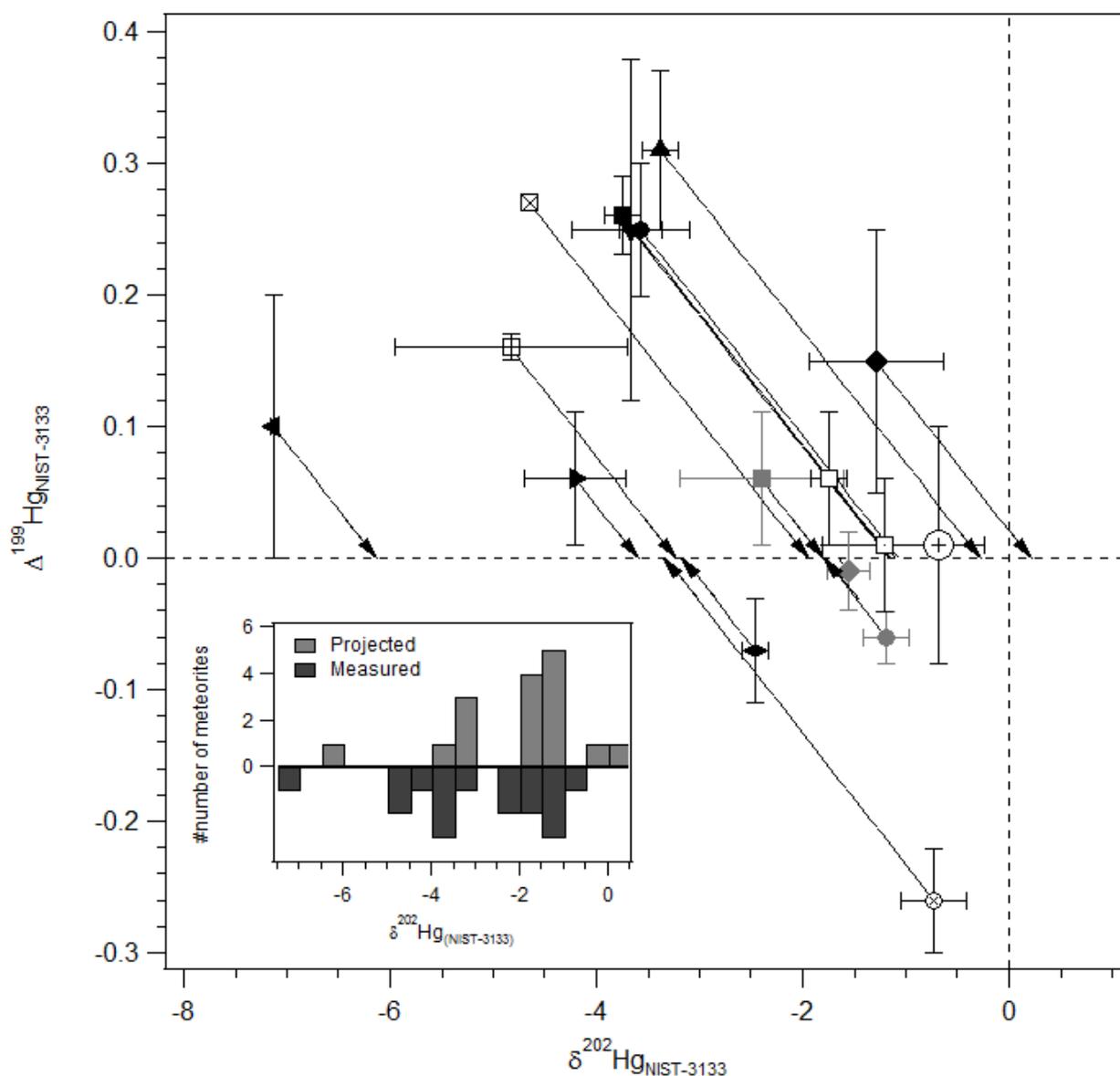

*Figure 9: Mass-independent (Δ199Hg) vs. mass-dependent (δ202Hg) fractionation of Hg isotopes. Symbols used are the same as in Fig. 1-5, the large crossed dot is the average composition of terrestrial crustal rocks (Blum et al., 2014). Inset shows a histogram of δ202Hg values, both measured and projected to Δ199Hg = 0 on a line with slope = -0.1 (see main text).*

## 5. Summary and Conclusions

We have measured the Mercury (Hg) abundance and isotopic composition in a suite of meteorites, including chondrites and achondrites. We use step-heating profiles to constrain the main carriers of Hg in meteorites, which include troilite, cinnabar, elemental Hg and perhaps some Hg-compounds (e.g. HgO, $HgSO_4$), which have yet to be observed *in situ*. We show that terrestrial Hg contamination is only a minor issue and likely contributes less than 10% of the total Hg in the meteorite samples analyzed in this study. However, meteorites that rested for a long time in a hot desert are more likely to have lost some of their Hg. All meteorites measured in this study show large variations in Hg abundance and mass-dependent fractionation of Hg isotopes, even within the same meteorite classes (and, in some cases, within the same meteorite), but abundance and isotopic composition do not correlate with each other. However, we find correlated mass-independent isotopic fractionation in meteorites, i.e., correlated enrichments and depletions of the odd-numbered Hg isotopes [199]Hg and [201]Hg. Most carbonaceous chondrites show enrichments in "vapor-phase Hg" (enriched in light and odd-numbered isotopes), while most non-carbonaceous chondrites and achondrites show either depletions or no variations relative to the terrestrial standard. We find no sign of nucleosynthetic [196]Hg, and conclude that the solar system must have been well-mixed with respect to supernova-derived Hg. We develop a series of simple models of thermally induced Hg migration in ordinary and carbonaceous chondritic asteroids, and show that migration can explain part of the observed abundances and isotopic fractionations, but additional processes, e.g., near-surface impact heating, are required to fully explain the observed isotopic and abundance variations. Finally, we show that the Earth's complement of Hg is likely to be predominantly derived from the chondritic source that also delivered water, nitrogen, carbon and the noble gases to Earth during late accretion. In that case, we would expect the Earth's mantle to be isotopically light in Hg compared to surface reservoirs.

## Acknowledgements

The authors thank Carl Agee, Alex Bewan, Ludovic Ferrière, Ingo Leya and Thomas Smith, Yves

Marrocchi, Ben McHenry, Mikhail Nazarov, François Robert, and Birger Schmitz for the contribution of meteorite samples for this study. We also thank Barbara Marie and Pierre-Yves Martin for their help in the DMA lab, Catherine Zimmermann for her (immense) patience and help with the acid extractions, Damien Cividini for his help with the ICP-MS analysis, and Guillaume Avice for his help with the melting of the basaltic glass sample using the noble gas line. MM also thanks Maria Schönbächler for helpful discussions and Greg deSouza for his kind help with setting-up of the MatLab model. We thank Kathrin Altwegg for informing us about the capability of Rosetta to detect Hg at 67P/Churyumov-Gerasimenko. Finally, we would like to thank Dante S. Lauretta, Joel D. Blum and an anonymous third reviewer for their constructive and insightful reviews. This work was supported by the European Research Council under the European Community's Seventh Framework Program (FP7/2010-2016 grant agreement no. 267255), and by the Swiss National Science Foundation (SNSF Ambizione grant to M. M., project no. PZ00P2_154874). This is CRPG publication No. XXXX.

**References**

Arlandini C., Käppeler F., Wisshak K., Gallino R., Lugaro M., Busso M., Straniero O. (1999) Neutron capture in low-mass asymptotic giant branch stars: cross sections and abundance signatures. Astrophys. J. 525:886-900.

Beer H., Macklin R. L., (1985) 198,199,200,201,202,204Hg(n,y) cross sections and the termination of s-process nucleosynthesis. Phys. Rev. C 32:738-755.

Bennet M. E., McSween H. Y. (1996) Revised model calculations for the thermal histories of ordinary chondrite parent bodies. Meteorit. Planet. Sci. 31:783-792.

Bergquist B. A., Blum J. D. (2007) Mass-dependent and mass-independent fractionation of Hg isotopes by photoreduction in aquatic systems. Science 318:417–420.

Blum J. D., Sherman L. S., Johnson M. W. (2014) Mercury Isotopes in Earth and Environmental Sciences. Annu. Rev. Earth Planet. Sci. 42:249-269.

Blum J. D., Bergquist B. A. (2007) Reporting the variations in the natural isotopic composition of mercury. Anal. Bioanal. Chem. 388:353–359.

Boss A. P., Keiser S. A. (2012) Supernova-triggered molecular cloud core collapse and the Rayleigh-Taylor fingers that polluted the solar nebula. Astrophys. J. Lett. 756:L9.

Bottke W. F., Vokrouhlický D., Rubincam D. P., Nesvorný D (2006) The Yarkovsky and YORP Effects: Implications for Asteroid Dynamics. Ann. Rev. Earth & Planet. Sci. 34:157-191.

Brennecka G. A., Borg L. E., Wadhwa M. (2013) Evidence for supernova injection into the solar nebula and the decoupling of r-process nucleosynthesis. Proc. Nation. Acad. Sci. 110:17241-17246.

Caillet Komorowski C., El Goresy A., Miyahara M., Boudouma O., Ma C. (2012) Discovery of Hg-Cu-bearing metal-sulfide assemblages in a primitive H-3 chondrite: Towards a new insight in early solar system processes. Earth Planet. Sci. Lett. 349-350:261-271.

Canil D., Crockford P. W., Rossin R., Telmer K. (2015) Mercury in some arc crustal rocks and mantle peridotites and relevance to the moderately volatile element budget of the Earth. Chem. Geol. 396:134-142.

Capaccioni F., Coradini A., Filacchione G., Erard S., Arnold G., Drossart P., De Sanctis M. C., Bockelee-Morvan D., Capria M. T., Tosi F., Leyrat C., Schmitt B., Quirico E., Cerroni P., Mennella V., Raponi A., Ciarniello M., McCord T., Moroz L., Palomba E., Ammannito E., Barucci M. A., Bellucci G., Benkhoff J., Bibring J. P., Blanco A., Blecka M., Carlson R., Carsenty U., Colangeli L., Combes M., Combi M., Crovisier J., Encrenaz T., Federico C.,

Fink U., Fonti S., Ip W. H., Irwin P., Jaumann R., Kuehrt E., Langevin Y., Magni G., Mottola S., Orofino V., Palumbo P., Piccioni G., Schade U., Taylor F., Tiphene D., Tozzi G. P., Beck P., Biver N., Bona L., Combe J.-Ph., Despan D., Flamini E., Fornasier S., Frigeri A., Grassi D., Gudipati M., Longobardo A., Markus K., Merlin F., Orosei R., Rinaldi G., Stephan K., Cartacci M., Cicchetti A., Giuppi S., Hello Y., Henry F., Jacquinod S., Noschese R., Peter G., Politi R., Reess J. M., Semery A. The organic-rich surface of comet 67P/Churyumov-Gerasimenko as seen by VIRTIS/Rosetta. Science 347, 6220.

Chen J., Hintelmann H., Dimock B. (2010) Chromatographic pre-concentration of Hg from dilute aqueous solutions for isotopic measurement by MC-ICP-MS. J. Anal. At. Spectrom. 25:1402-1409.

Chen J., Hintelmann H., Feng X., Dimock B. (2012) Unusual fractionation of both odd and even mercury isotopes in precipitation from Peterborough, ON, Canada. Geochim. Cosmochim. Acta 90:33-46.

Ciesla F. J., Davison T. M., Collins G. S., O'Brien D. P. (2013) Thermal consequences of impacts in the early solar system. Meteorit. Planet. Sci. 48:2559-2576.

Demers J. D., Blum J. D., and Zak D. R. (2013) Mercury isotopes in a forested ecosystem: Implications for air-surface exchange dynamics and the global mercury cycle. Glob. Biogeochem. Cycl. 27:222-238.

Dreibus G., Palme H., Spettel B., Zipfel J., Wanke H. (1995). Sulfur and selenium in chondritic meteorites. Meteoritics 30:439-445.

Ebihara M., Kumar P., Bhattacharya S. K. (1998) 196Hg/202Hg ratio and Hg content in meteorites and terrestrial standard rocks:; A RNAA study. Lunar Planet. Sci. Conf. 29, Abstract 1727.

Estrade N., Carignan J., Sonke J. E., Donard O.F.X. (2009) Mercury isotope fractionation during liquid-vapor evaporation experiments. Geochim. Cosmochim. Acta 73:2693-2711.

Estrade N., Carignan J., Donard O.F.X. (2010) Isotope tracing of atmospheric mercury sources in an urban area of northeastern France. Environ. Sci. Technol. 44:6062-6067.

Fischer-Gödde M., Becker H. (2012) Osmium isotope and highly siderophile element constraints on ages and nature of meteoritic components in ancient lunar impact rocks. Geochim. Cosmochim. Acta 77:135-156.

Ghosh S., Schauble E. A., Lacrampe Couloume G., Blum J. D., Bergquist B. A. (2013) Estimation of nuclear volume dependent fractionation of mercury isotopes in equilibrium liquid-vapor evaporation experiments. Chem. Geol. 336:5-12.

Gladstone G. R., Hurley D. M., Retherford K. D., Feldman P. D., Pryor W. R., Chaufray J.-Y., Versteeg M., Greathouse T. K., Steffl A. J., Throop H., Parker J. W., Kaufmann D. E., Egan A. F., Davis M. W., Slater D. C., Mukherjee J., Miles P. F., Hendrix A. R., Colaprete A., Stern S. A. (2010) LRO-LAMP Observations of the LCROSS Impact Plume. Science 330:472-476.

Gounelle M., Spurný P., Bland P. A., 2006. The orbit and atmospheric trajectory of the Orgueil meteorite from historical records. Meteorit. Planet. Sci. 41:135-150.

Grevesse N. (1970) Solar and meteoritic abundances of mercury. Geochim. Cosmochim. Acta 34:1129-1130.

Grevesse N., Scott P., Asplund M., Jacques Sauval A. (2015) The elemental composition of the Sun III. The heavy elements Cu to Th. Astron. Astroph. 573:A27.

Haack H., Michelsen R., Stober G., Keuer D., Singer W., Williams I. (2011) CM chondrites from comets? - New constraints from the orbit of the Maribo CM chondrite fall. Form. First Solids Solar Syst., Abstract 9100.

Herndon J. M., Rowe M. W., Larson E. E., Watson D. E. (1975) Origin of magnetite and pyrrhotite in carbonaceous chondrites. Nature 253:516-518.

Hewins R. H., Bourot-Denise M., Zanda B., Leroux H., Barrat J.-A., Humayun M., Göpel C., Greenwood R. C., Franchi I. A., Pont S., Lorand J.-P., Cournède C., Gattacceca J., Rochette P., Kuga M., Marrocchi Y., Marty B. (2014) The Paris meteorite, the least altered CM chondrite so far. Geochim. Cosmochim. Acta 124:190-222.

Jaschek C., Jaschek M. (1995) The Behavior of Chemical Elements in Stars. Cambridge University Press, Cambridge, United Kingdom.

Huss G. R., Lewis R. S., and Hemkin S., 1996. The "normal planetary" noble gas component in primitive chondrites: Compositions, carrier, and metamorphic history. Geochim. Cosmochim. Acta 60:3311-3340.

Klaue, B., Kesler, S. E. and Blum, J. D. (2000). Investigation of the natural fractionation of stable mercury isotopes by ICPMS. In International Conference on Heavy Metals in the Environment. Ann Arbor, MI, USA.

Kumar P., Ebihara M., Bhattacharya S. K. (2001) $^{196}$Hg/$^{202}$Hg ratio and Hg content in meteorites and terrestrial standard rocks: A RNAA study. Geochem. J. 35:101-116.

Lauretta D. S., Devouard B., Buseck P. R. (1999) The cosmochemical behavior of mercury. Earth Planet. Sci. Lett. 171:35-47.

Lauretta D. S., Klaue B., Blum J. D., Buseck P. R. (2001) Mercury abundances and isotopic

compositions in the Murchison (CM) and Allende (CV) carbonaceous chondrites. Geochim. Cosmochim. Acta 65:2807-2818.

Marie B., Marin L., Martin P.-Y., Gulon T., Carignan J., Cloquet C. (2015) Determination of Mercury in One Hundred and Sixteen Geological and Environmental Reference Materials Using a Direct Mercury Analyser. Geostandards and Geoanalytical Res. 39:71-86.

Marrocchi Y., Gounelle M., Blanchard I., Caste F., Kearsley A. T. (2014) The Paris CM chondrite: Secondary minerals and asteroidal processing. Meteorit. Planet. Sci. 49:1232-1249.

Marty B. (2012) The origins and concentrations of water, carbon, nitrogen and noble gases on Earth. Earth Planet. Sci. Lett. 313-314:56-66.

Mason R. P. and Sheu G. R. (2002). Role of the ocean in the global mercury cycle. Glob. Biogeochem. Cycl. 16:40-41.

McDonough W. F., Sun S.-S. (1995) The composition of the Earth. Chemical Geology 120:223-253.

Mead C., Lyons J. R., Johnson T. M., Anbar A. D. (2013) Unique Hg Stable Isotope Signatures of Compact Fluorescent Lamp-Sourced Hg. Environ. Sci. Technol. 47:2542-2547.

Meier M. M. M. (2014) Are CM, CI chondrites samples from Jupiter Family Comets? 77[th] Ann. Meet. Meteorit. Soc., Abstract 5009.

Meier M. M. M., Cloquet C., Marty B. (2015a) Towards Mercury (Hg) cosmochemistry: variable contributions of supernova-derived Hg, or mass-independent fractionation by photodegradation? Lun. Planet. Sci. Conf. 46, Abstract 1101.

Meier M. M. M., Cloquet C., Marty B. (2015b) Making Sense of Mercury Isotopic and Abundance Variations in Meteorites. 78th Ann. Meet. Meteorit. Soc., abstract 5021.

Miyamoto M., Fujii N., Takeda H. (1981) Ordinary chondrite parent body – An internal heating model. Proc. Lunar Planet. Sci. 12B:1145-1152.

Palme H., Beer H. (1993) Abundances of the elements in the solar system. In Landolt Börnstein Group VI, Astronomy and Astrophyics, Vol 2A, ed. H. H. Voigt, pp. 196-221. Springer Verlag, Berlin, Germany.

Palme H., Kurat G., Brandstätter F., Burghele A., Huth J., Spettel B. and Wlotzka F. (1985) An unusual chondritic fragment from the Allende meteorite. Lunar Planet. Sci. Conf. XVI, pp. 645-646 (abstract).

Rauscher T., Heger A., Hoffman R. D., Woosley S. E. (2002) Nucleosynthesis in massive stars with improved nuclear and stellar physics. Astrophys. J. 576:323-348.

Reed G.W. (1999) Don't drink the water. Meteorit. Planet. Sci. 34:809-811.

Rose-Weston L., Brenan J. M., Fei Y., Secco R. A., Frost D. J. (2009) Effect of pressure, temperature, and oxygen fugacity on the metal-silicate partitioning of Te, Se, and S: Implications for earth differentiation. Geochim. Cosmochim. Acta 73:4598-4615.

Schaefer L. and Fegley B. (2010) Volatile element chemistry during metamorphism of ordinary chondritic material and some of its implications for the composition of asteroids. Icarus 205:483-496.

Seargent D. A. J. (1990) The Murchison meteorite: Circumstances of its fall. Meteoritics 25:341-342.

Sharma R. C., Chang Y. A. (1993) The Hg-S (Mercury-Sulfur) System. J. Phase Equilib. 14:100-109.

Sherman L. S., Blum J. D., Johnson K. P., Keeler G. J., Barres J. A., Douglas T. A. (2010) Mass-independent fractionation of mercury isotopes in Arctic snow driven by sunlight. Nat. Geosci. 3:173–177.

Smith R. S., Wiederhold J. G., Jew A. D., Brown G. E. Jr., Bourdon B., Kretzschmal R. (2014) Small-scale studies of roasted ore wast reveal extreme ranges of stable mercury isotope signatures. Geochim. Cosmochim. Acta 137:1-17.

Sugiura N., Fujiya W. (2014) Correlated accretion ages and $\varepsilon^{54}$Cr of meteorite parent bodies and the evolution of the solar nebula. Meteorit. Planet. Sci. 49:772-787.

Veverka J., Thomas P., Harch A., Clark B., Bell J.F. III, Carcich B., Joseph J., Chapman C., Merline W., Robinson M., Malin M., McFadden L.A., Murchie S., Hawkins S. E. III, Farquhar R., Izenberg N., Cheng A. (1997) NEAR's Flyby of 253 Mathilde: Images of a C Asteroid. Science 278:2109-2114.

Wasson J.T. (1972) Formation of Ordinary Chondrites. Rev. Geophys. Space Phys. 10:711-759.

Wasson J. T. and Kallemeyn G. W. (1988) Composition of Chondrites. Philosophical Transactions of the Royal Society of London, Series A, Mathematical and Physical Sciences, 325:535-544.

Weidenschilling S.J. (2011) Initial sizes of planetesimals and accretion of the asteroids. Icarus 214:671-684.

Wheelock M. M., Keil K., Floss C., Taylor G. J., Crozag G. (1994) REE geochemistry of oldhamite-dominated clasts from the Norton County aubrite: Igneous origin of oldhamite. Geochim. Cosmochim. Acta 58:449-458.

Wiederhold J. G., Schönbächler M. (2015) Mercury concentrations and Hg isotope compositions of chondrites and eucrites. Lunar Planet. Sci. Conf. 46, Abstract 1841.

Wlotzka F. (2005) Cr spinel and chromite as petrogenetic indicators in ordinary chondrites: Equilibration temperatures of petrologic types 3.7 to 6. Meteorit. Planet. Sci. 40:1673-1702.

Wombacher F., Rehkämper M., Mezger K., Münker C. (2003). Stable isotope compositions of cadmium in geological materials and meteorites determined by multiple-collector ICPMS. Geochim. Cosmochim. Acta 67:4639-4654.

Wombacher F., Rehkämper M., Mezger K., Bischoff A., Münker C. (2008) Cadmium stable isotope cosmochemistry. Geochim. Cosmochim. Acta 72:646-667.

Yin R., Feng X., Meng B. (2013) Stable Mercury Isotope Variation in Rice Plants (Oryza sativa L.) from the Wanshan Mercury Mining District, SW China. Environ. Sci. Technol. 47:2238-2245.

Yokoyama T., Fukami Y., Nagai Y., Nakamoto T. (2014) Volatility control of isotope heterogeneity in the early solar system. Lunar Planet. Sci. Conf. 45 Abstract 2588.

Yund R. A., Hall H. T. (1968) The miscibility gap between FeS and Fe1-xS. Mat. Res. Bull. 3:779-784.

Zambardi T., Sonke J.E., Toutain J.P., Sortinob F., Shinoharac H., (2009). Mercury emissions and stable isotopic compositions at Vulcano Island (Italy). Earth Planet. Sci. Lett. 277:236–243.

**Table 1: All meteorite Hg data, averages**

| Sample | Type | Source | Hg (ppb) | $\delta^{202}$Hg | $\Delta^{196}$Hg | $\Delta^{199}$Hg | $\Delta^{200}$Hg | $\Delta^{201}$Hg | $\Delta^{204}$Hg | Sessions (n) |
|---|---|---|---|---|---|---|---|---|---|---|
| *Carbonaceous chondrites* | | | | | | | | | | |
| Orgueil | CI1 | MNHN | 13600±5200 | -3.74±0.18 | 0.50±1.14 | 0.26±0.03 | 0.06±0.04 | 0.16±0.07 | -0.03±0.20 | I, II, III (20) |
| Tagish Lake | C2-ungr. | COM1 | 71±1 | -3.57±0.20 | 1.24±1.94 | 0.25±0.05 | 0.06±0.03 | 0.14±0.02 | 0.00±0.09 | V (4) |
| Murchison | CM2 | CRPG | 2570±50 | -1.29±0.65 | - | 0.15±0.10 | 0.01±0.01 | 0.06±0.04 | - | I (3) |
| Jbilet Winselwan | CM2 | CRPG | 50 | - | - | - | - | - | - | - |
| Dhofar 1432 | CR2 | COM1 | 12±2 | - | - | - | - | - | - | - |
| Paris | CM2.7 | CRPG | 11800±2500 | -3.38±0.18 | - | 0.31±0.06 | 0.01±0.08 | 0.14±0.05 | - | I, II (9) |
| Kainsaz | CO3.2 | ASM | 267±29 | -7.13±0.03 | 0.44±0.94 | 0.10±0.10 | 0.03±0.11 | 0.07±0.12 | -0.02±0.09 | IV (3) |
| Allende | CV3 | COM1 | 39±13 | -3.66±0.57 | -1.43±1.98 | 0.25±0.13 | 0.07±0.04 | 0.08±0.22 | 0.04±0.01 | V (2) |
| Karoonda | CK4 | SAM | 401±73 | -2.46±0.12 | 0.05±0.24 | -0.07±0.04 | 0.02±0.04 | -0.09±0.09 | -0.04±0.01 | IV (3) |
| Bencubbin silicates | CB | WAM | 202±122 | -4.20±0.49 | 0.82±1.55 | 0.06±0.05 | -0.02±0.04 | 0.06±0.06 | 0.00±0.04 | V (6) |
| *Ordinary, enstatite and Rumuruti chondrites* | | | | | | | | | | |
| NWA 7323 | LL3 | CRPG | 19±6 | - | - | - | - | - | - | - |
| Mezö-Madaras | L3.7 | NHMV | 3150±520 | -1.19±0.23 | 0.44±0.35 | -0.06±0.02 | -0.03±0.04 | -0.10±0.08 | -0.40±0.24 | I, II (8) |
| Mezö-Madaras, heated* | L3.7 | NHMV | ca. ~1500 | -1.55±0.20 | 0.36±0.43 | -0.01±0.03 | 0.00±0.01 | -0.02±0.01 | -0.02±0.01 | V (4) |
| Saratov | L4 | CRPG | 18±1 | - | - | - | - | - | - | - |
| Ghubara | L5 | LU | 44 | - | - | - | - | - | - | - |
| Mocs | L6 | NHMV | 965±140 | -2.40±0.79 | - | 0.06±0.05 | 0.01±0.03 | -0.01±0.07 | - | I (6) |
| Tieschitz | H3.6 | UBA | 24 | -6.00±0.89 | - | - | - | - | - | V (2) |
| Sahara 97072 | EH3 | COM1 | 34 | -4.39±1.17 | - | - | - | - | - | - |
| NWA 753 | R3.9 | COM1 | 8±1 | - | - | - | - | - | - | - |
| *Achondrites* | | | | | | | | | | |

| Sample | Type | Source | Hg (ppb) | $\delta^{202}$Hg | $\Delta^{196}$Hg | $\Delta^{199}$Hg | $\Delta^{200}$Hg | $\Delta^{201}$Hg | $\Delta^{204}$Hg | Sessions (n) |
|---|---|---|---|---|---|---|---|---|---|---|
| Millbillillie | Eucrite | WAM | 28±2 | -0.73±0.31 | 1.31±1.52 | -0.26±0.04 | -0.08±0.07 | -0.13±0.04 | -0.20±0.06 | III (3) |
| Norton County, bulk | Aubrite | UNMA | 94±1 | -1.75±0.18 | - | 0.06±0.05 | 0.02±0.05 | -0.03±0.06 | - | I (3) |
| Norton County, enstatite | Aubrite | UNMA | 15±9 | - | - | - | - | - | - | - |
| Norton County, dark | Aubrite | UNMA | 627±20 | -1.20±0.60 | - | 0.01±0.05 | 0.01±0.03 | -0.01±0.07 | - | I (3) |
| Almahata Sitta | Ureilite | COM2 | 203±1 | -4.83±1.12 | - | 0.16±0.01 | 0.02±0.05 | -0.03±0.09 | - | V (2) |
| Mundrabilla troilite | IAB-ungr | UBE | 8 | -4.65 | - | 0.27 | -0.12 | 0.25 | 0.05 | V (1) |
| *Terrestrial standards (averages)* | | | | | | | | | | |
| NIST-3133 | Reference | NIST | - | 0.00±0.03 | - | 0.00±0.01 | 0.00±0.01 | 0.00±0.01 | - | I (97) |
| | | | | 0.00±0.06 | 0.07±1.04 | 0.00±0.02 | 0.00±0.02 | 0.00±0.03 | -0.01±0.12 | II (48) |
| | | | | -0.01±0.13 | -0.01±0.24 | 0.00±0.01 | 0.00±0.01 | 0.00±0.01 | 0.00±0.01 | III (70) |
| | | | | 0.01±0.11 | 0.00±0.77 | 0.00±0.01 | 0.00±0.01 | 0.00±0.01 | 0.00±0.02 | V (36) |
| UM-Almadén | Standard | CRPG | - | -0.32±0.22 | - | -0.03±0.03 | -0.01±0.02 | -0.07±0.02 | - | I (13) |
| | | | | -0.48±0.41 | 0.75±2.26 | -0.04±0.05 | 0.00±0.07 | -0.02±0.08 | 0.17±0.23 | II (9) |
| | | | | -0.37±0.14 | 0.52±0.88 | -0.01±0.02 | 0.01±0.02 | -0.02±0.02 | 0.01±0.03 | III (16) |
| F65A | Standard | CRPG | - | -2.91±0.41 | - | 0.10±0.04 | 0.04±0.02 | 0.04±0.03 | - | I (6) |
| | | | | -3.47±0.35 | 0.95±2.59 | 0.09±0.03 | 0.00±0.05 | 0.04±0.05 | -0.16±0.13 | II (4) |
| RL24 | Standard | CRPG | - | 2.11±0.41 | - | -0.03±0.03 | 0.00±0.02 | -0.04±0.04 | - | I (8) |
| | | | | 2.38±0.10 | -1.31±1.79 | 0.01±0.11 | 0.01±0.03 | 0.00±0.02 | 0.00±0.05 | II (3) |

*Table 1: Averages of Hg elemental concentration (determined with the DMA) and isotopic composition for all meteorites analyzed during this study. Uncertainties correspond to two standard deviations of individual measurements for concentration data, and two standard errors of the mean for isotopic compositions. If no uncertainty is given, only a single measurement / cycle was done. The last column indicates the sessions (extractions) in which a particular meteorite was analyzed for its isotopic composition, and (in brackets) the number of replicate ICP-MS measurements over all sessions (each measurement involves 40 cycles). Note that $^{196}$Hg and $^{204}$Hg were not measured in session I. For some meteorites, only concentration data are given when no isotopic analysis was attempted. For other meteorites, only mass-dependent isotopic fractionations are given as the Hg abundance in the sample solution was too low to allow the reliable analysis of mass-independent isotopic anomalies, while for others, only the high-abundance isotopes 199-201 were analyzed. For academic sample source abbreviations, see main text (commercial sources: COM1, Eric Twelker / meteoritemarket.com; COM2, Gabriele and Stephan Decker / meteorite-shop.de). \*Mezö-Madaras after pre-heating to 150°C, see main text.*

**Table 2: Step heating of selected meteorite samples**

| Sample | Type | Mass (g) | 150°C | 200°C | 250°C | 300°C | 350°C | 400°C | 450°C | 650°C | Total |
|---|---|---|---|---|---|---|---|---|---|---|---|
| Orgueil | CI1 | 0.0744 | 817 (0.082) | 225 (0.023) | 900 (0.090) | 1270 (0.127) | 2260 (0.226) | 1800 (0.180) | 1650 (0.165) | 1060 (0.106) | 9980 (1) |
| Orgueil* | CI1 | 0.0250 | 1500 (0.143) | 261 (0.025) | 1200 (0.114) | 1750 (0.167) | 2480 (0.236) | 1580 (0.150) | 1200 (0.114) | 564 (0.054) | 10500 (1) |
| Orgueil*, atm | CI1 | 0.0211 | 180 (0.700) | 22.2 (0.0863) | 34.8 (0.135) | 10.4 (0.0404) | 3.5 (0.0136) | 2.09 (0.00813) | 1.24 (0.00482) | 2.94 (0.0114) | 257 (1) |
| Orgueil*, atm | CI1 | 0.0175 | 396 (0.628) | 58.4 (0.093) | 130 (0.206) | 25.0 (0.040) | 6.2 (0.010) | 4.7 (0.007) | 4.2 (0.007) | 6.9 (0.011) | 631 (1) |
| Allende | CV3 | 0.141 | 11.9 (0.385) | 1.8 (0.058) | 6.7 (0.217) | 4.1 (0.133) | 1.2 (0.039) | 1.2 (0.039) | 1.5 (0.049) | 2.5 (0.081) | 31 (1) |
| Karoonda | CK4 | 0.150 | 171 (0.454) | 22.9 (0.061) | 73.2 (0.194) | 47.7 (0.127) | 22.9 (0.061) | 3.6 (0.010) | 16.5 (0.044) | 19.1 (0.051) | 377 (1) |
| Bencubbin | CB | 0.638 | 29.2 (0.206) | 10.6 (0.075) | 27.5 (0.194) | 7.2 (0.051) | 1.4 (0.010) | 3.3 (0.023) | 15.3 (0.108) | 47.2 (0.332) | 142 (1) |
| Mocs | L6 | 0.0990 | 454 (0.432) | 52.3 (0.050) | 203 (0.193) | 161 (0.153) | 37.3 (0.036) | 27.1 (0.026) | 51.4 (0.049) | 65.9 (0.063) | 1050 (1) |
| Mezö-Madaras | L3.7 | 0.117 | 2120 (0.642) | 172 (0.052) | 583 (0.177) | 170 (0.052) | 58.4 (0.018) | 52.0 (0.016) | 65.4 (0.020) | 76.2 (0.023) | 3300 (1) |
| Millbillillie | Eucrite | 0.268 | 25.8 (0.806) | 1.5 (0.047) | 2.1 (0.066) | 0.9 (0.028) | 0.5 (0.016) | 0.4 (0.013) | 0.3 (0.009) | 0.5 (0.016) | 32.0 (1) |
| Norton County** | Aubrite | 0.123 | 326 (0.446) | 56.9 (0.078) | 132 (0.181) | 70.5 (0.096) | 22.1 (0.030) | 13 (0.018) | 17.5 (0.024) | 93.3 (0.128) | 731 (1) |
| Mundrabilla*** | IAB | 0.0232 | 10.8 (0.246) | 3.2 (0.073) | 2.8 (0.064) | 2.2 (0.050) | 1.7 (0.039) | 1.7 (0.039) | 15.9 (0.362) | 5.6 (0.128) | 43.9 (1) |

*Table 2: Hg concentrations (in ppb or ng/g) released at different temeprature steps. Values in brackets are normalized to the total abundance of Hg in a meteorite (see also Fig. 3). \*These samples were pre-heated to 100°C during 22 hours on a hot plate. Samples labeled with "atm" were additionally exposed to the atmosphere for three months before pre-heating and analysis. \*\*Dark phase only. \*\*\*Troilite inclusion.*

**Table 3: Possible carrier phases of Hg in different meteorites**

| Step (°C) | OR | AL | KA | BE | MO | MM | MI | NC | MU | Possible Hg carriers | Release T (°C) |
|---|---|---|---|---|---|---|---|---|---|---|---|
| 150 | √ | √√ | √√ | √ | √√ | √√ | √√ | √√ | √ | Troilite → pyrrhotite? Other (Millbillillie)? | ca. 140 |
| 200 | × | × | × | × | × | × | × | × | × | Adsorbed terrestrial Hg | ca. 195 |
| 250 | √ | √ | √ | √ | √ | √ | √ | √ | × | Hg-compounds (HgS, HgO, HgSe) | ca. 230-270 |
| 300 | √ | × | × | × | √ | × | × | √? | × | Unknown | - |
| 350 | √√ | × | √? | × | × | × | × | × | × | Boiling point of native Hg; $\alpha$HgS → $\beta$HgS | 357, 345 |
| 400 | √ | × | × | √ | × | × | √? | × | × | Unknown | - |
| 450 | √ | √ | √ | √ | √ | √ | × | √ | √√ | Troilite | ? |
| 650 | √ | √ | √ | √√ | √ | √ | √ | √ | √ | HgO, HgSO$_4$ | 550, 600 |

*Table 3: Possible Hg carriers matching peaks observed in the thermal release profile of Hg (Fig. 3) in different meteorites. Meteorite abbreviations: OR = Orgueil; AL = Allende; KA = Karoonda; BE = Bencubbin; MO = Mocs; MM = Mezö-Madaras; MI = Millbillillie; NC = Norton County; MU = Mundrabilla. A (√) symbol stands for the likely presence of the carrier, given that a local peak (compared to neighboring temperature steps) is observed, while a (×) symbol stands for a local minimum or the non-observation of a local peak. Maximum fractional release per meteorite is indicated by a (√√) symbol. In the last column, the likely release temperatures of the respective carriers (mostly from Lauretta et al., 2001) are given. The temperature of the troilite-pyrrhotite transition (140°C) is given by Yund and Hall (1968).*

## Supplementary Material

### Parent body heating model calculations

The input parameters for each of the four models presented in the main text are given in table S1.

**Table S1: Input parameters for the Hg redistribution models (Fig. 6 and 7)**

| Model | Chondrite | Radius (km) | Density (kg/m$^3$) | Porosity (%) | $T_{init}$-$t_0$ (Ma) | $T_{amb}$, $T_{initial}$ (K) | H at $T_{amb}$ (km) |
|-------|-----------|-------------|--------------------|--------------|----------------------|------------------------------|---------------------|
| OC85 | Ordinary | 85 | 3600 | 2.5 | 2.1 | 190 | 18 |
| CC20 | Carbonaceous | 20 | 1500 | 50 | 4.0 | 170 | 169 |

*For the parameter study, the radius was varied between 10, 20, 40, 80 and 160 km, density and porosity were co-varied between 2850/5%, 2700/10%, 2400/20%, and 1500/50%, the formation time was varied between 3, 3.5, and 4 Ma. The ambient temperature was kept at 170 K.*

We first divide the asteroid in 1000 concentric shells of equal mass (to capture the processes near the surface at higher resolution). We then use the Miyamoto et al. (1981) model – originally developed for ordinary chondrites – to determine the temperature in each of the 1000 shells as a function of time (the first time-step is at 10'000 a, with a factor of $2^{(1/100)}$ between each following time-step, and 1563 time-steps in total, corresponding to 500 Ma). To simulate the evolution of Hg concentration and isotopes with time (and temperature), we first set the initial Hg concentration to 260 ppb (= Hg$_{BS3}$). Then, the vapor pressure of Hg (in Pa) at each time-step, and for each shell is evaluated as a function of temperature using the Clausius-Clapeyron relation for Hg:

$$P_{(T)} = 36.38 \times e^{\,(61400\,/\,8.134)\,\times\,(1\,/\,373\,-\,1\,/\,T)} \tag{1}$$

We then assume that the vapor phase Hg is present at the Hg vapor pressure using the ideal gas equation, assuming a porosity of 2.5% for a compacted L-chondritic asteroid (Bennet & McSween, 1996) and 10% for the carbonaceous chondrite model. Note that we do include gravitational compaction of pore spaces since even the core pressures of the modeled asteroids (calculated as $2/3\pi G\rho^2 R^2$) are still well below the compressive strength of meteorites. The porosity volume of each shell (V) is filled with the necessary amount of Hg (in mol) to reach the vapor pressure $P_{(T)}$ calculated with (1) above. The local temperature T to calculate the vapor pressure is derived from the Miyamoto et al. (1981) temperature model (R = 8.134 J mol$^{-1}$ K$^{-1}$; gas constant). For this, we use

the ideal gas law:

$$n = P_{(T)} \times V / (R \times T) \qquad\qquad (2)$$

The amount of Hg needed for this (the "vapor fraction") is subtracted from (and not allowed to exceed) the total amount of Hg present in each shell at the beginning of each time-step. Then, the vapor Hg is summed up over the entire asteroid and re-distributed among all shells according to gravity and temperature. By doing this, we implicitly assume that the gaseous Hg is free to migrate in a completely connected interior pore network, forming an internal asteroidal atmosphere. However, no Hg is lost to space in this model. The pressure in this internal atmosphere is given by:

$$P_{atm} = P_0 \times e^{(-z/H)} \qquad\qquad (3)$$

Here, $P_0$ is the pressure at a reference level, z is the height above the reference level and H the scale height. The scale height can also be expressed as:

$$H = R \times T / g \qquad\qquad (4)$$

where g is the local gravitational acceleration. As shown in Table S1, the scaling height of the internal atmosphere (at ambient temperature and gravitational acceleration near the surface) is significantly smaller than the asteroid radius for the 85 km OC, while it is also significantly smaller than radius for the 20 km CC model. Replacing g in (4) with $G \times M / z^2$ and using it the resulting expression to replace H in (3) we get:

$$P_{atm} = P_0 \times e^{(-G \times M / (z \times R \times T))} \qquad\qquad (5)$$

We then use again the ideal gas law to determine the amount of Hg to be found in each shell:

$$n = P_{atm} \times V / (R \times T) \qquad\qquad (6)$$

We use this relationship to calculate the relative abundance of Hg in all shells (i.e., the value of $P_0$ does not need to be defined since we are only interested in relative abundances). We then redistribute the vapor fraction Hg (in each time-step) proportionally back into the shells.

As the temperature rises (initially) through time, the fraction of vapor phase Hg relative to total Hg increases in most of the shells (except the outer-most ones which always stay cold), until a temperature is reached – in some shells – where all Hg is in the vapor phase. The Hg redistribution mechanism then leads to almost complete re-deposition of Hg in the cold outer crust (typically the outermost few km, with the highest values reached near the surface).

In order to track the evolution of mass-dependent and mass-independent isotope fractionation, we first determine the mass-dependent isotopic fractionation from vaporization at each time-step and for each shell (at its given temperature T) using the temperature-fractionation-relationship given by Estrade et al. (2009):

$$\delta^{202}Hg = 9.44 - 1.4 \times (10^6 / T^2) \qquad \text{if T < 385 K} \qquad (4)$$
$$\delta^{202}Hg = 0 \qquad \text{if T > 385 K}$$

For each time-step, all the vapor-phase Hg within the asteroid is assumed to equilibrate, i.e., we calculate the isotopic composition of the mixture of all vapor-phase Hg from all shells at a given time-step. The redistributed Hg then adopts this isotopic composition, and is mixed (in a mass-balanced way) with the residual (condensed) Hg in each of the shells. The intitial $\delta^{202}Hg$ is set to (defined as) zero.

Finally, the mass-independent fractionation of Hg is simply set by the empirical relationship $\Delta^{199}Hg = -0.1 \times \delta^{202}Hg$ (Blum et al., 2014; note that in that publication, the slope is given as +0.1, but it should actually be -0.1 because enrichments in heavy isotopes (= larger $\delta^{202}Hg$ values) go together with a depletion (not an enrichment!) of odd-numbered isotopes in the residual phase as shown, e.g., by the experiments of Estrade et al., 2009).